\documentclass[journal]{IEEEtran}

\usepackage{ifthen}

\newboolean{printTODO}
\newboolean{printAbstract}

\setboolean{printTODO}{false}
\setboolean{printAbstract}{true}

\makeatletter
\def\markboth#1#2{\def\leftmark{\@IEEEcompsoconly{\sffamily}\MakeUppercase{\protect#1}}%
\def\rightmark{\@IEEEcompsoconly{\sffamily}\MakeUppercase{\protect#2}}}
\makeatother

\usepackage[american]{babel}
\usepackage[utf8]{inputenc} 

\usepackage{grffile}

\usepackage{etex}

\usepackage[hidelinks]{hyperref}

\ifCLASSINFOpdf
  \usepackage[pdftex]{graphicx}
	\usepackage{epstopdf}
  \graphicspath{{graphics/}}
  \DeclareGraphicsExtensions{.eps,.pdf,.jpeg,.jpg,.png}
\else
  \usepackage[dvips]{graphicx}
  \graphicspath{{graphics/}}
  \DeclareGraphicsExtensions{.eps}
\fi

\usepackage[cmex10]{amsmath}
\usepackage{array}

\usepackage{eqparbox}

\usepackage[caption=false,font=footnotesize]{subfig}
\usepackage{subfig}

\usepackage{stfloats}

\usepackage{url}

\ifthenelse{\boolean{printTODO}}
{
	\usepackage{tikz}
	\usepackage[colorinlistoftodos,textsize=scriptsize]{todonotes}
	\setlength{\marginparwidth}{1.3cm}
}
{
	\usepackage[disable,colorinlistoftodos]{todonotes}
}

\usepackage[backend=biber,style=ieee,sorting=none,doi=true,isbn=false,url=false,minbibnames=1,maxbibnames=1,mincitenames=1,maxcitenames=1]{biblatex}
\usepackage[babel]{csquotes}
\uspunctuation

\usepackage[alsoload=astro, alsoload=hep]{siunitx}
\usepackage{placeins} %

\usepackage{isotope}

\usepackage{xspace}
\makeatletter
\let\@autoref=\autoref
\renewcommand*{\autoref}[2][]{\ifthenelse{\equal{#1}{}}{\@autoref{#2}}{\hyperref[#1]{\begin{NoHyper}\@autoref{#2}~\subref{#1}\end{NoHyper}}}\xspace}
\makeatother

\usepackage{epstopdf}

\usepackage{verbatim}

\usepackage{booktabs}
\usepackage{xcolor,colortbl}
\usepackage{tabularx}
\newcolumntype{C}[1]{>{\centering\arraybackslash}m{#1}}
\newcolumntype{R}[1]{>{\raggedleft\arraybackslash}m{#1}}

\usepackage{multirow}

\DeclareFieldFormat{sentencecase}{\MakeSentenceCase{#1}}
\renewbibmacro*{title}{%
  \ifthenelse{\iffieldundef{title}\AND\iffieldundef{subtitle}}
  {}
  {\ifthenelse{\ifentrytype{article}\OR\ifentrytype{inbook}%
      \OR\ifentrytype{incollection}\OR\ifentrytype{inproceedings}%
      \OR\ifentrytype{inreference}}
    {\printtext[title]{%
        \printfield[titlecase]{title}%
        \setunit{\subtitlepunct}%
        \printfield[titlecase]{subtitle}}}%
    {\printtext[title]{%
        \printfield[titlecase]{title}%
        \setunit{\subtitlepunct}%
        \printfield[titlecase]{subtitle}}}%
    \newunit}%
  \printfield{titleaddon}
}

\usepackage{tikz}

\usepackage{adjustbox}

\newcommand{\tikzsubfig}[3]
{
	\subfloat{
		\label{#3}
		\begin{tikzpicture}
		\node[anchor=south west,inner sep=0] (image) at (0,0) {\includegraphics[#2]{#1}};
		\begin{scope}[x={(image.south east)},y={(image.north west)}]
		\node[font=\small] at (0.06, 0.06) {\subref{#3}};
		\end{scope}
		\end{tikzpicture}
	}
}

\begin{document}

\newcommand{\myTitle}{Initial Measurements with the PETsys TOFPET2 ASIC Evaluation Kit and a Characterization of the ASIC TDC}

\title{{\LARGE \myTitle}}

\author{David~Schug$^{1}$,
        Vanessa~Nadig$^{1}$,
        Bjoern~Weissler$^{1}$,
        Pierre~Gebhardt$^{1}$,
        and~Volkmar~Schulz$^{1}$%
\thanks{$^1$Department of Physics of Molecular Imaging Systems, Institute for Experimental Molecular Imaging, RWTH Aachen University, Aachen, Germany}%
\thanks{This project is funded by the German Research Foundation (DFG), project number SCHU 2973/2-1.}
\thanks{This project has received funding from the European Union’s Horizon 2020 research and innovation programme under grant agreement No 667211 \protect\includegraphics[height=0.75em]{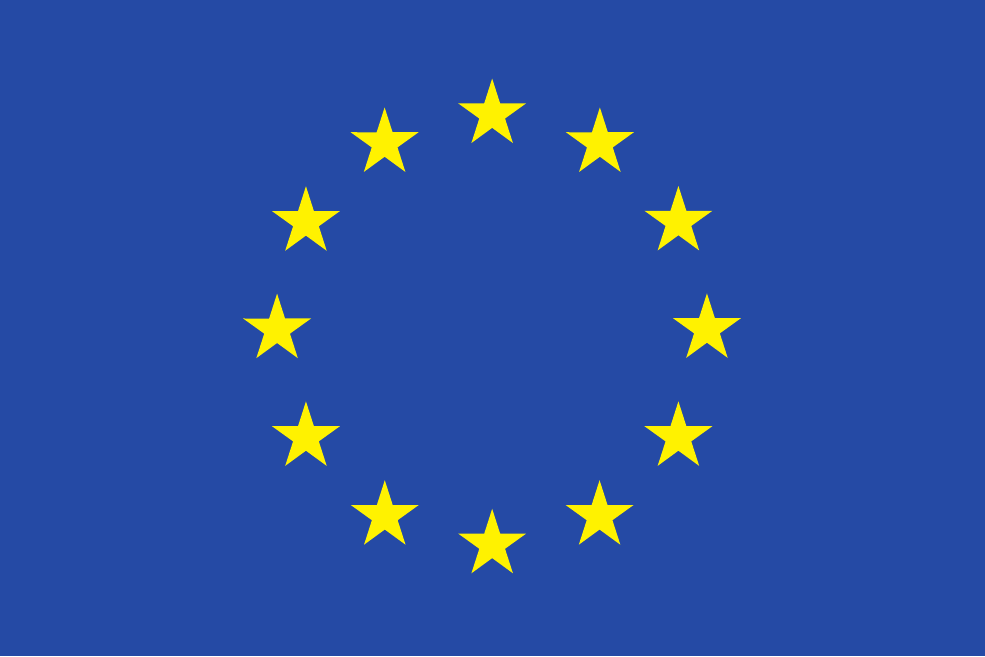}.}
}

\markboth{submitted 15/08/2018 to: IEEE TRPMS - Speicial Issue PSMR2018 - 7th conference on PET/MR and SPECT/MR}%
{Schug \MakeLowercase{\textit{et al.}}: \myTitle}

\maketitle

\ifthenelse{\boolean{printAbstract}}
{
\begin{abstract}
	
For a first characterization, we used the two KETEK-PM3325-WB SiPMs each equipped with a \SI{3x3x5}{mm} LYSO scintillation crystal provided with the PETsys TOFPET2 ASIC Evaluation Kit. We changed the lower of two discriminator thresholds (D\_T1) in the timing branch from vth\_t1 = 5-30. The overvoltage was varied in a range of  \SIrange{1.25}{7.25}{\volt}. The ambient temperature was kept at \SI{16}{\celsius}. For all measurements, we performed an energy calibration including a correction for saturation. We evaluated the energy resolution, the coincidence resolving time (CRT) and the coincidence rate. At an overvoltage of \SI{6}{\volt}, we obtained an energy resolution of about 10\%~FWHM, a CRT of approximately \SI{210}{\ps}~FWHM and \SI{400}{\ps}~FWTM, the coincidence rate showed only small variations of about 5\%.
To investigate the influence of the ambient temperature, it was varied between \SIrange{12}{20}{\celsius}.
At \SI{12}{\celsius} and an overvoltage of \SI{6.5}{\volt}, a CRT of approx. \SI{195}{\ps}~FWHM and an energy resolution of about 9.5\%~FWHM could be measured.
Observed satellite peaks in the time difference spectra were investigated in more detail.
We could show that the location of the satellite peaks is correlated with a programmable delay element in the trigger circuit.

\end{abstract}
}{}

\section{Introduction}

\IEEEPARstart{P}{ositron} emission tomography (PET) is an imaging technique based on the annihilation of a positron with an electron and the resulting emission of two \SI{511}{\keV} gamma photons in opposite directions. For clinical usage, tracers -- biologically active molecules labeled with positron-emitting radionuclides -- are injected into the body.
The emitted gamma photons have to be detected in coincidence by a PET scanner.
One assumes that the line connecting the locations of the two gamma interactions with the PET detector (line of response, LOR) includes the point of annihilation and thus the location of the tracer.
By measuring many LORs and using a tomographic image reconstruction, the spatial distribution of the tracer in the patient's body is computed \cite{surti2016advances,vandenberghe2016RecentDevelopments,moses2007RecentAdvances}.
PET is an essential tool in the diagnosis and staging of cancer as well as, e.g., in measuring cardiac perfusion and the assessment of Alzheimer's disease
 \cite{glass2013PETmammography,weber2003PETLungCancer,nakazato2013MyocardialPerfusion,marcus2014BrainPET,rinne2010AlzheimerAssessment,barshalom2003clinicalperformance}.

Typically, PET detectors are arranged in a ring geometry and consist of scintillators, which convert the energy of the annihilation gamma photon into optical photons, photo detectors, readout and digitization electronics as well as mechanics, housing and cooling elements.
Performance parameters of a PET detector are the spatial resolution, the timing resolution and the energy resolution.
If the timing resolution of the PET detectors is sufficient, the difference in arrival times of the two gamma interactions can be used to localize the annihilation event along the LOR resulting in a non-uniform probability distribution of the position of the annihilation event along the LOR.
This is called time-of-flight PET (TOF-PET).
TOF-PET systems were developed in the late 1980s and require to measure the time difference of the two gamma interactions (coincidence resolving time, CRT) with a timing resolution in the order of a few hundred picoseconds \cite{vandenberghe2016RecentDevelopments,surti2016advances}.

In most state-of-the-art PET systems, lutetium(-yttrium) oxy-orthosilicate (L(Y)SO) scintillators are used to convert the gamma photons into optical photons due to their favorable properties such as high light output ($\sim\SI{25000}{\text{ph.}\per\MeV}$), short decay time ($\sim\SI{40}{\ns}$) and fast rise time ($\sim\SI{70}{\ps}$), which result in excellent CRT values \cite{surti2016advances,bisogni2015solidstatephotodetectors,gundacker2016measurement}. %
In addition to a fast scintillator to improve the timing resolution in PET systems \cite{moses2007RecentAdvances}, it is important to employ fast photodetectors to convert the optical photons into an electrical signal.
Today, Silicon photomultipliers (SiPMs) are used for this purpose due to their compactness and lower voltage requirements compared to photomultiplier tubes (PMTs).
Furthermore, SiPMs can be operated in a magnetic field why they are the photosensor of choice for applications that integrate a PET into an MRI  \cite{vandenberghe2016RecentDevelopments,surti2016advances,moses2007RecentAdvances}.
An SiPM consists of several thousands of Geiger-mode cells, so-called single-photon avalanche diodes (SPAD), each generating a very similar analog signal upon breakdown \cite{bisogni2015solidstatephotodetectors}.
This breakdown is ideally caused by an impinging optical photon but can also be thermally induced in the semiconductor material.
This material and temperature dependent noise rate is typically in the order of \SIrange{50}{100}{\kilo\text{cps}\per\square\mm} at room temperature.
In a single analog SiPM, also referred to as pixel or channel, several thousand SPADs are connected in parallel.
The total signal of the analog SiPM is proportional to the number of detected optical photons and the rising edge of the signal contains timing information in the order of a few tens of picoseconds \cite{bisogni2015solidstatephotodetectors,KETEKProductDataSheet,gulinatti2005SPADresolution,cova1989SPADresolution,cova1981SPADresolution}.

\begin{figure}[tb]
  \centering
  \includegraphics[height=0.92\columnwidth]{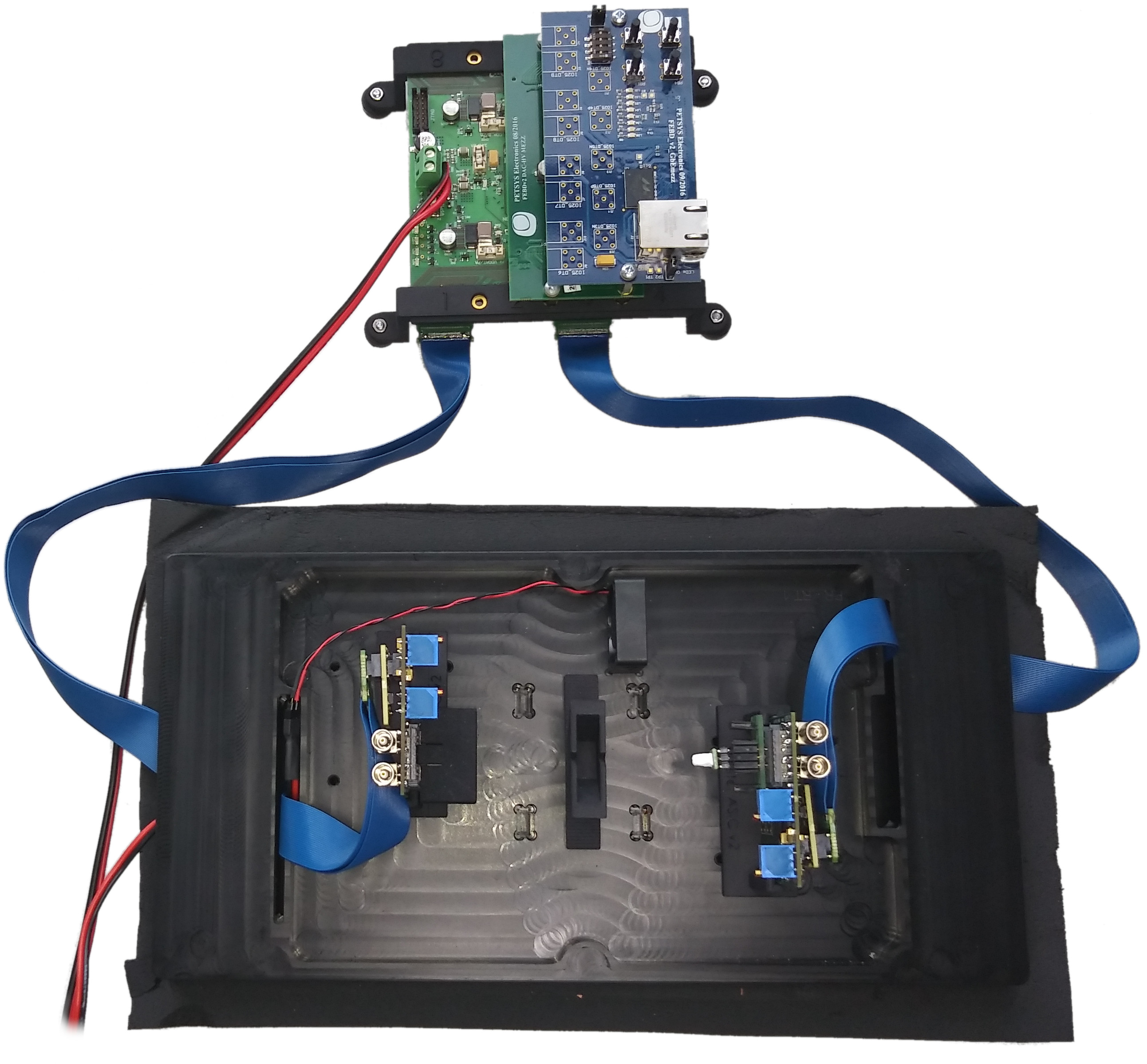}
  \caption{Coincidence measurements with the TOFPET2 setup showing a KETEK PM3325-WB-A0 SiPMs with \SI{3x3x5}{\mm} LYSO crystals connected to TOFPET2 ASIC evaluation boards shown in the bottom of the picture. The ASIC evaluation boards are connected to the data acquisition board (FEB/D) which houses high-voltage regulators and a Gigabit Ethernet communication interface.}
  \label{fig:setup_opened_with_crystal}
\end{figure}

TOF-PET systems that utilize fast scintillators coupled to analog SiPMs often employ an application-specific integrated circuit (ASIC) to digitize the analog SiPM signals with the goal to precisely measure the energy and timestamp of the gamma interaction with the scintillator material.
In general, there's a wide variety of ASICs designed for various applications in medical imaging or nuclear physics \cite{basiladze2016ASICreview}. The present work concentrates on ASICs specifically designed for TOF-PET applications.
These ASICs must be capable of digitizing the timestamp and the energy with a high precision in order not to significantly deteriorate the information provided with the analog SiPM signal.
The time\-stamp is often acquired using a comparator set to a very low threshold to trigger on the rising edge of the SiPM signal.
The timing resolution is thus influenced by the characteristics of the scintillator, the SiPM, the ASIC employed and the electrical interconnections of these components.
There are benchtop experiments which can achieve coincidence time resolutions below \SI{100}{\ps} \cite{nemallapudi2015sub100ps,sarasola2017FlexToTvsNINO}.
On system level, several TOF-PET-ASIC-development projects aim to allow for a CRT better than \SI{300}{\ps} or even as low as \SI{200}{\ps} \cite{rolo2012TOFPETASIC,chen2014DedicatedReadoutASIC,bugalho2017ExperimentalResultsTOFPET2}.
Energy can be measured using a second timestamp on the falling edge of the SiPM signal.
From the difference between the two timestamps, the time-over-threshold (ToT) value can be deducted which contains information about the integral of signal \cite{chen2014DedicatedReadoutASIC,orita2017CurrentToTASIC}.
An integration of the analog SiPM signal can be realized using capacitors to store the energy of the signal \cite{corsi2009ASICdevelopment}.

The recently developed TOF-PET ASICs are designed in standard CMOS technology with a varying number of input channels ranging from 8 channels \cite{corsi2009ASICdevelopment} over 16 channels \cite{fischer2009PETA,shen2012STiC}, 32 channels \cite{fleury2017PETIROC2A}, 36 channels \cite{sacco2013PETA4} and 64 channels \cite{rolo2012TOFPETASIC,PETsysTOFPET1DataSheet,chen2014DedicatedReadoutASIC,ahmad2015Triroc,ahmad2016Triroc}.
Of these previously mentioned TOF-PET ASICs, for example the PETA series \cite{fischer2009PETA,fischer2006MultiChannelReadout,piemonte2012PerformancePETA3ASIC,sacco2013PETA4,sacco2015PETA5,schug2017PETA5} as well as the Weeroc series are using an integration method for the energy measurement \cite{ahmad2015Triroc,ahmad2016Triroc,fleury2017PETIROC2A}. %
A PET insert for a 7T MRI scanner (MADPET4) using the TOFPET v1 ASIC \cite{omidvari2017MADPET4}, a tri-modal PET/MR/EEG scanner (TRIMAGE) using the Triroc ASIC \cite{ahmad2015Triroc,ahmad2016Triroc,sportelli2016TRIMAGEInitialResults} and a multi-modal setup combining TOF-PET with ultrasound endoscopy (EndoToFPET-US) using the STiC ASIC \cite{shen2012STiC,chen2014DedicatedReadoutASIC} are only three of many examples for the application of TOF-PET ASICs.

A promising candidate for high-performance TOF-PET applications is the 64-channel TOFPET2 ASIC recently developed and presented by PETsys Electronics S.A.
Designed in standard CMOS technology, this ASIC provides independent energy and time digitization for each of the 64 input channels.
The digital part of the ASIC is clocked with a frequency of \SI{200}{\MHz}  \cite{PETsysFlyer,francesco2016TOFPET2}.
Compared to its prior version, time binning was optimized from \SI{50}{\ps} to \SI{30}{\ps} \cite{PETsysTOFPET1DataSheet,PETsysFlyer} and the trigger scheme was changed from a dual mixed-mode to a three-threshold readout scheme (for detailed description see section \ref{sec:triggergeneration}).
Moreover, the energy measurement method can be switched between a ToT mode and an integration mode (called "qdc" mode by PETsys) \cite{rolo2013TOFPETASIC, rolo2012TOFPETASIC}.

PETsys distributes a so-called TOFPET2 ASIC evaluation kit, which provides the electronics and software required to test the ASIC in a small bench-top setup.
It includes two ASIC evaluation boards and a data acquisition board to synchronize and configure the ASICs as well as capture the digitized data which can be connected via a network link to a computer running a control and data acquisition software.
The kit is pre-equipped with two single KETEK SiPMs coupled to small LYSO crystals enabling the user to easily acquire measurements with two SiPMs in coincidence.
Larger SiPM arrays and scintillator matrices may be connected to the setup and would allow the investigation of the ASIC performance with two TOFPET2 ASICs, with 64 channels each, in coincidence.
In this work, we present initial results obtained with the provided single-channel setup, report the performance results obtained and artifacts observed in the time difference measurements and investigate their cause.

\section{Materials}

\begin{figure}[tb]
  \centering
  \includegraphics[height=0.92\columnwidth]{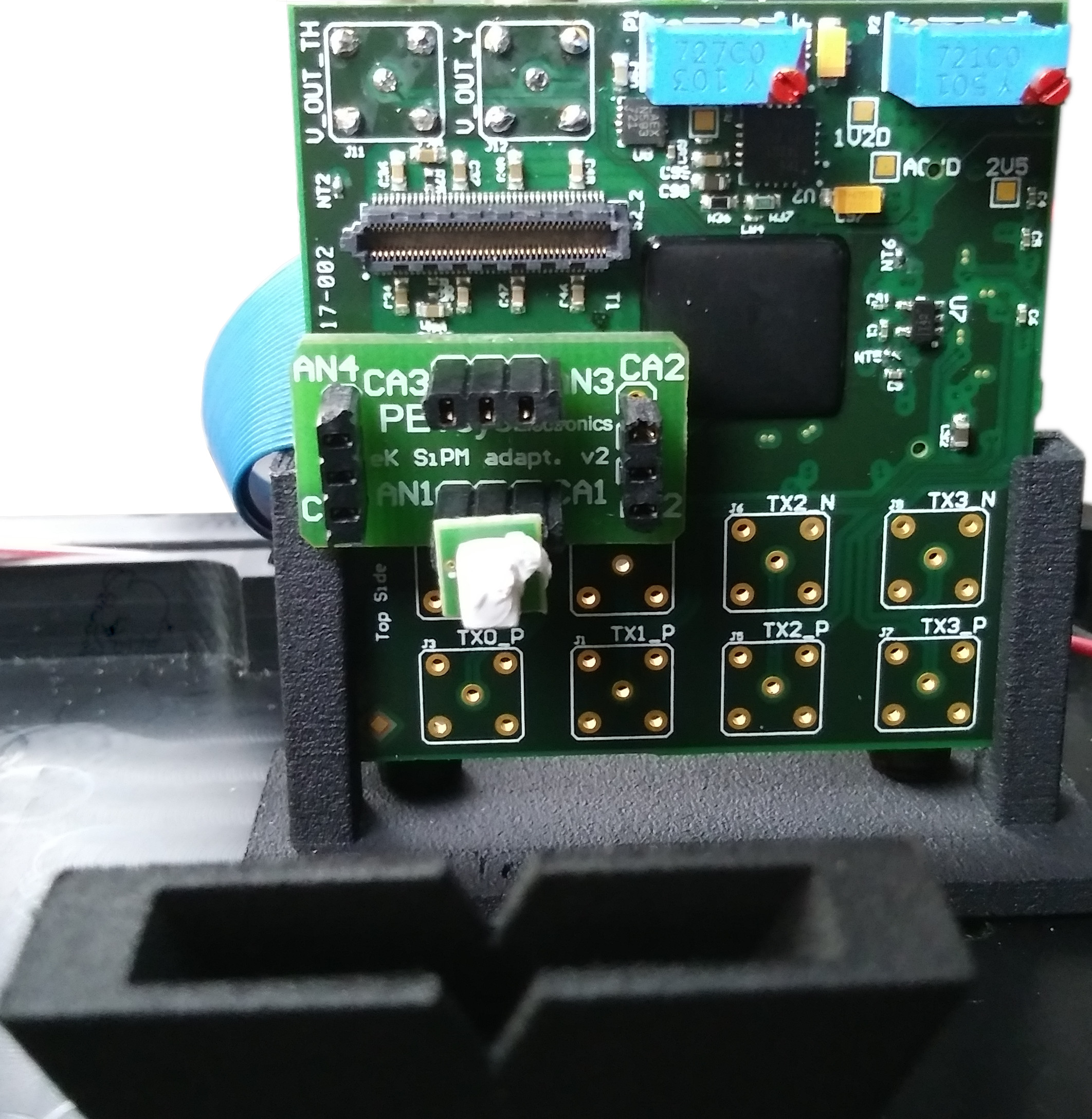}
  \caption{KETEK PM3325-WB-A0 SiPM with \SI{3x3x5}{\mm} LYSO crystal connected to TOFPET2 ASIC board.}
  \label{fig:KETEK_SiPM_on_ASIC}
\end{figure}

In this work, we used the TOFPET2 ASIC evaluation kit, which provides two TOFPET2 ASIC evaluation boards.
Each board houses a TOFPET2 ASIC with 64 analog input channels \cite{PETsysTOFPET2DataSheet,PETsysFlyer,TOFPET2EKitHardwareGuide,TOFPET2EKitSoftwareGuide}.
To both boards, a single KETEK-PM3325-WB SiPM \cite{KETEKProductDataSheet} was connected, each equipped with a \SI{3x3x5}{\mm} LYSO scintillation crystal (\autoref{fig:KETEK_SiPM_on_ASIC}).
The breakdown voltage of the used SiPMs was assumed to be \SI{27}{\volt} according to the technical documentation \cite{KETEKProductDataSheet}.
This value was verified with an error of \SI{0.2}{\volt} for the combination of SiPM and crystal by a linear regression of the \SI{511}{\keV} peak position.
We used a \isotope[22]{Na} point source with an active diameter of \SI{0.5}{\mm} and an activity of about \SI{8.}{\mega\becquerel} located in the center between the two detector elements.
The holders of the TOFPET2 ASIC test boards were mounted in the closest position of the three available positioning options per ASIC-evaluation-board holder, which leads to a distance of about \SI{65}{\mm} between crystal front faces.
For data acquisition, the two boads were connected to a FEB/D board with a FPGA and powered with \SI{12}{\volt} DC using flexible cables.
The FEB/D board generates the central clocking signal and distributes it to the ASIC evaluation boards.
To generate the programmable positive SiPM bias voltages, a high-voltage(HV)-DAC mezzanine board was mounted to the FEB/D board.
The effective bias voltage $V_\text{bias}$ applied to the SiPM is about \SI{0.75}{\volt} lower compared to the set voltage according to the PETsys manual \cite{TOFPET2EKitSoftwareGuide}.
All overvoltages reported in this work were corrected for this value.
The value of \SI{0.75}{\volt} was verified with an error of \SI{0.1}{\volt} by probing the voltage applied to the SiPM with a multimeter for multiple channels.
The FEB/D board was connected to a computer via a Gigabit Ethernet link for communication.
All firmware and data acquisition software to set global as well as channel-specific ASIC configuration parameters, e.g. trigger configuration and threshold settings, were provided by PETsys. 
Furthermore, a calibration script and tools to process the captured raw data to hit data containing channel IDs, timestamps and energy values were shipped with the evaluation kit.
A light-tight aluminium box, in which the setup was placed, was attached to a peltier cooling element controlled by a PID controller to stabilize the ambient temperature in the box which was measured using a Pt100 temperature probe.
Pictures of the setup are shown in \autoref{fig:setup_opened_with_crystal} and \autoref{fig:KETEK_SiPM_on_ASIC}.
More details on the hardware can be found in \cite{TOFPET2EKitHardwareGuide}.

\subsection*{Details on Trigger and Timestamp Generation}
\label{sec:triggergeneration} 

To avoid confusion, we use the naming convention for components, thresholds and signals that PETsys uses in their ASIC datasheet documentation \cite{PETsysTOFPET2DataSheet}.

\begin{figure}[tb]
    \centering
    \includegraphics[width=1.00\columnwidth]{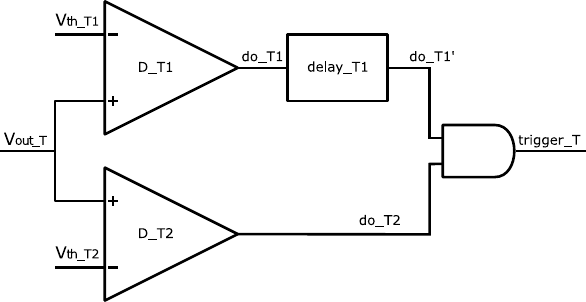}
    \caption{Sketch of the TOFPET2 ASIC trigger scheme used in the timing branch. Adapted from \cite{PETsysTOFPET2DataSheet}}
    \label{fig:fig03}
\end{figure}

The input voltage $V_\mathsf{out\_T}$ that is evaluated by the timing-trigger circuit is generated by a transimpedance amplifier from the SiPM input signal and should already be corrected for a baseline.
The TOFPPET2 ASIC uses a two-level analog trigger scheme in the timing branch to discriminate noise events without introducing dead time (\autoref{fig:fig03}).
The first discriminator $\mathsf{D\_T1}$ is set to a low voltage threshold $V_\mathsf{th\_T1}$ and its output $\mathsf{do\_T1}$ is hold back by a configurable delay element $\mathsf{delay\_T1}$ resulting in a delayed signal $\mathsf{do\_T1'}$ which is fed into one input of an $\mathsf{AND}$ gate.
A second discriminator $\mathsf{D\_T2}$ is set to a value $V_\mathsf{th\_T2}$ corresponding to a higher voltage value as $V_\mathsf{th\_T1}$.
The output of that second discriminator $\mathsf{do\_T2}$ is not delayed when routed to the second input of the $\mathsf{AND}$ gate.
If both inputs of the $\mathsf{AND}$ gate are active, the $\mathsf{trigger\_T}$ signal is generated, which is routed to the time-to-digital converter and which defines the time stamp of the light-pulse event recorded by that ASIC channel.
For a rising voltage signal, the $\mathsf{do\_T2}$ signal will always be generated at a later point in time $t_\mathsf{T2}$ than the $\mathsf{do\_T1}$ signal at $t_\mathsf{T1}$.
To invert the order of the discriminator signals at the $\mathsf{AND}$ gate, the $\mathsf{delay\_T1}$ element should be set to a value that is higher than the highest expected value of the time difference between $t_\mathsf{T2}$ and $t_\mathsf{T1}$  which then ensures that $\mathsf{do\_T1'}$ is always reaching the $\mathsf{AND}$ gate after $\mathsf{do\_T2}$.
In the separated energy branch, the discriminator threshold $\mathsf{D\_E}$ can be set to reject noise events.

\section{Methods}

The ambient temperature in the box was set to \SI{16}{\celsius} for the measurements performed to investigate the performance as a function of the  $V_\mathsf{th\_T1}$ threshold and overvoltage.
According to the Software user guide \cite{TOFPET2EKitSoftwareGuide}, the LSB of the discriminator $\mathsf{D\_T1}$ was set to $\mathsf{disc\_lsb\_t1} = \num{60}$, which should result in approximately \SI{2.5}{\milli\volt} per DAC step \cite{PETsysTOFPET2DataSheet}.
The discriminator thresholds $\mathsf{D\_T2}$ and $\mathsf{D\_E}$ operate on a different lsb scale and were kept at their default settings $\mathsf{disc\_lsb\_t2} = \num{48}$ and $\mathsf{disc\_lsb\_e} = \num{40}$, which correspond to about \SI{15.0}{\milli\volt} and  \SI{20.0}{\milli\volt} per DAC step, respectively \cite{PETsysTOFPET2DataSheet,PETsys_priv_comm}.
We used the energy integrating "qdc" mode.
The calibration routine, which is part of the software tools provided with the evaluation kit, was run at an overvoltage of $V_\text{OV} = \SI{4.25}{\volt}$ with the discriminator thresholds set to the following default values of DAC steps above the baseline $\mathsf{vth\_t1} = 20$, $\mathsf{vth\_t2} = 20$ and $\mathsf{vth\_e} = 15$ \cite{TOFPET2EKitSoftwareGuide}.
By default, we programmed the length of the $\mathsf{delay\_T1}$ element by setting $\mathsf{fe\_delay} = \num{14}$ which corresponds to a theoretical delay length of $t_\mathsf{delay\_T1} = \SI{5.8}{\ns}$ \cite{PETsys_priv_comm}.
As the theoretical delay value $t_\mathsf{delay\_T1}$ reported by PETsys is not the result of a calibration of the $\mathsf{delay\_T1}$ element, the actual value might be slightly different.
For the main measurement series, we changed the discriminator threshold $\mathsf{vth\_t1} = \numrange{5}{30}$ in steps of \num{5} and the overvoltage  $V_\text{OV} = \SIrange{1.25}{7.25}{\volt}$ in steps of \SI{0.25}{\volt} and acquired data during \SI{60}{\second}.
At an overvoltage of $V_\text{OV} = \SI{4.25}{\volt}$, the discriminator threshold range set should correspond to a signal height in photo electrons (or SPAD breakdowns) of about \numrange{0.5}{3}.
The obtained ASIC calibration was used for all measurements of this series.
We developed an analysis software which works with the corrected hits (called "singles" by PETsys) provided by the PETsys software "\texttt{convert\_raw\_to\_singles}".

For each of the \num{150} measurements, an individual energy calibration per detector was performed and then used during data processing.
The \SI{511}{\keV} and \SI{1275}{\keV} peaks in the energy value histogram (\autoref{fig:crystal_0_energy_value_spectrum}) were identified by calculating and subtracting the background \cite{Ryan1988396, Morhac1997113, Burgess1983431} and performing a peak search using a Markov chain algorithm \cite{Morhac2000108}.
A Gaussian was fitted to the peaks.
A simple saturation model was applied and parameterized using the positions of the two energy peaks neglecting energy offsets or optical crosstalk (cf. eq. \eqref{eq:saturationmodel}).
The result is the saturation-corrected energy $E$, determined by the energy value $e$, the energy factor $c$ in \si{\keV} per unit of $e$ and the saturation parameter  $s$ in units of $e$ (\autoref{fig:crystal_0_energy_spectrum}).
These parameters were determined numerically using the fitting routine of ROOT.
No noise pedestal or energy offset besides the constant correction determined during the calibration routine was calculated.
Singles with an energy of \SIrange{400}{700}{\keV} were considered for the coincidence search using a coincidence window of \SI{35}{\nano\second} to be sensitive to any problems that might lead to timestamp errors up to several clock cycles.

\begin{equation}
E = c \cdot s \cdot \ln\left(\frac{1}{1-\frac{e}{s}}\right)
\label{eq:saturationmodel}
\end{equation}

For all measurements, the time difference between the two detectors and the energy spectrum of coincident events were evaluated.
The energy resolution and CRT as full width at half maximum (FWHM) were determined by fitting a Gaussian to a defined range of the respective spectrum (details of the fitting methods are described in \cite{schug2016petperformance}, see \autoref{fig:coincident_energy_spectrum}).
To probe the tail behavior of the time difference distribution, we determined the full width at tenth maximum (FWTM).

To investigate the performance as a function of temperature, a further measurement series was conducted.
For these measurements the ambient temperature was varied between \SIrange{12}{20}{\celsius} in steps of \SI{2}{\celsius} using a constant setting for the $V_\mathsf{th\_T1}$ threshold of $\mathsf{vth\_t1}= \num{30}$ and the same overvoltage range as for the first measurement series.
Data was processed in the same way as for the measurement series described above.

To qualitatively investigate the cause of observed satellite peaks (two smaller peaks in the time difference histogram located symmetrically around the main peak) in the time difference spectrum further we conducted four measurements using the non-default values for $\mathsf{fe\_delay}$ of $\num{11}$, $\num{12}$ and $\num{13}$ which correspond to a delay length $t_\mathsf{delay\_T1}$ of \SIlist{12.9;0.39;2.95}{\ns}, respectively additionally to the default value \cite{PETsys_priv_comm}.

Then, we set the discriminator threshold $\mathsf{vth\_t1}$ to the minimum value of 1, which should result in the highest fraction of coincidence events with an abnormal time difference, and varied the overvoltage for values of the delay element $t_\mathsf{delay\_T1}$ of \SIlist{2.95;5.8;12.9}{\ns}.
For these measurements, the calibration method was repeated before each of the three delay settings applied.
In the time difference spectrum we determined the difference of the satellite peak positions with respect to the central peak by fitting a Gaussian to all three peaks.
This measurement series was conducted to reveal a dependence of the peak position with the slope of the SiPM signal.
The peak position was evaluated by fitting a Gaussian to all peaks appearing in the time difference spectrum and then computing the distance of the satellite peaks to the main peak.

\section{Results}

\begin{figure}[tb]
  \centering
  \tikzsubfig{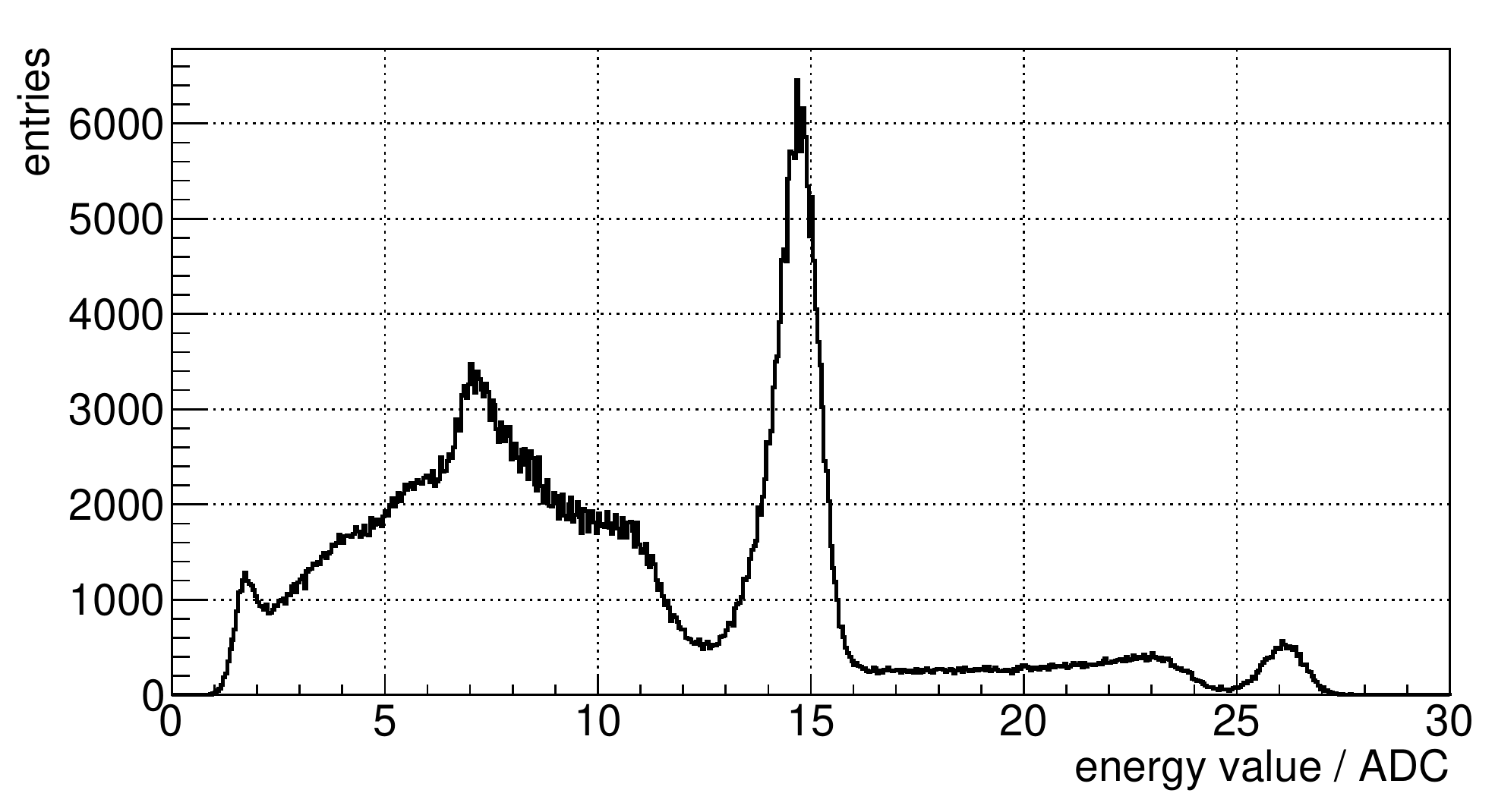}{width=1.00\columnwidth}{fig:crystal_0_energy_value_spectrum}
  \\
  \tikzsubfig{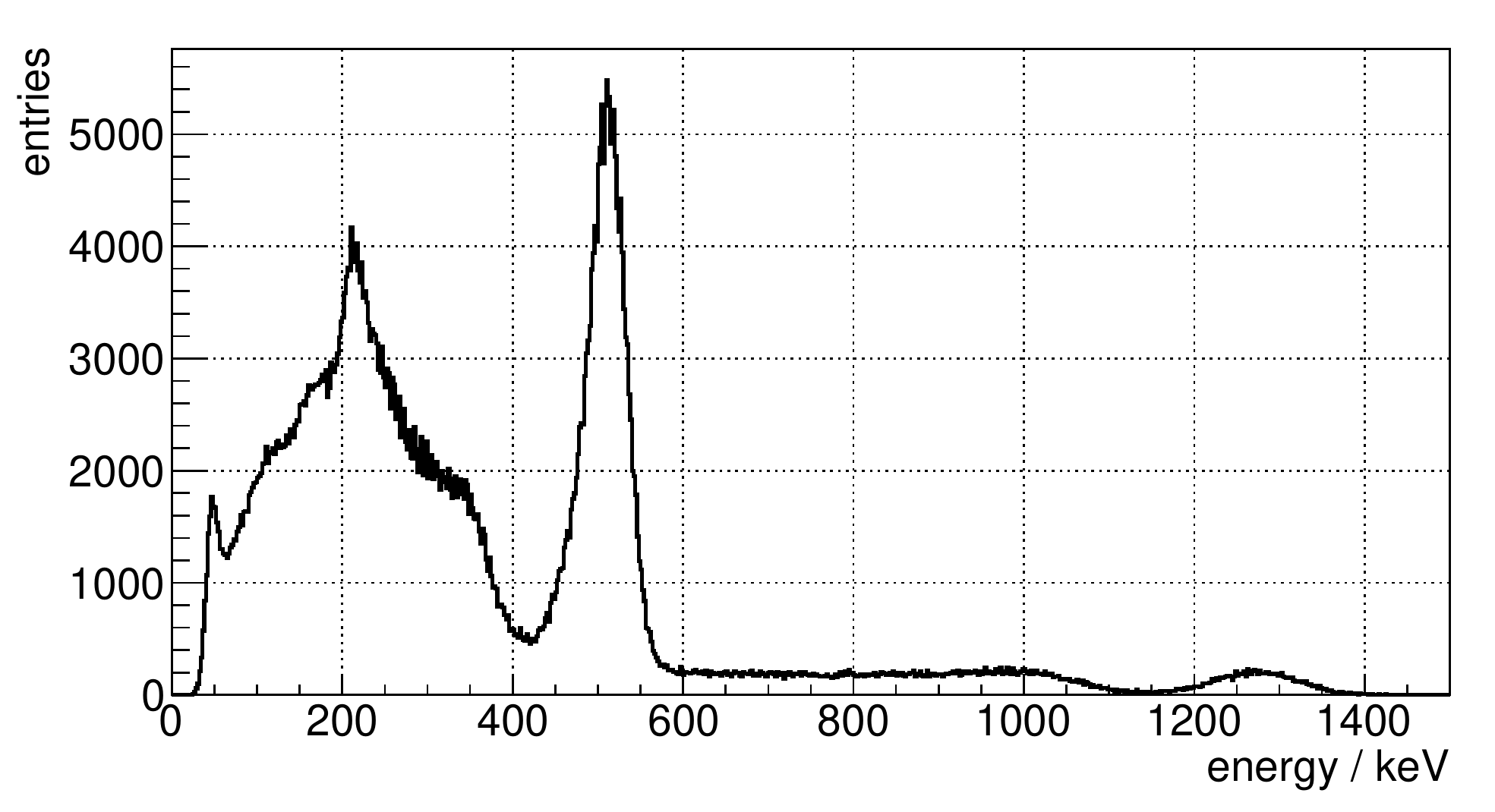}{width=1.00\columnwidth}{fig:crystal_0_energy_spectrum}
  \\
  \tikzsubfig{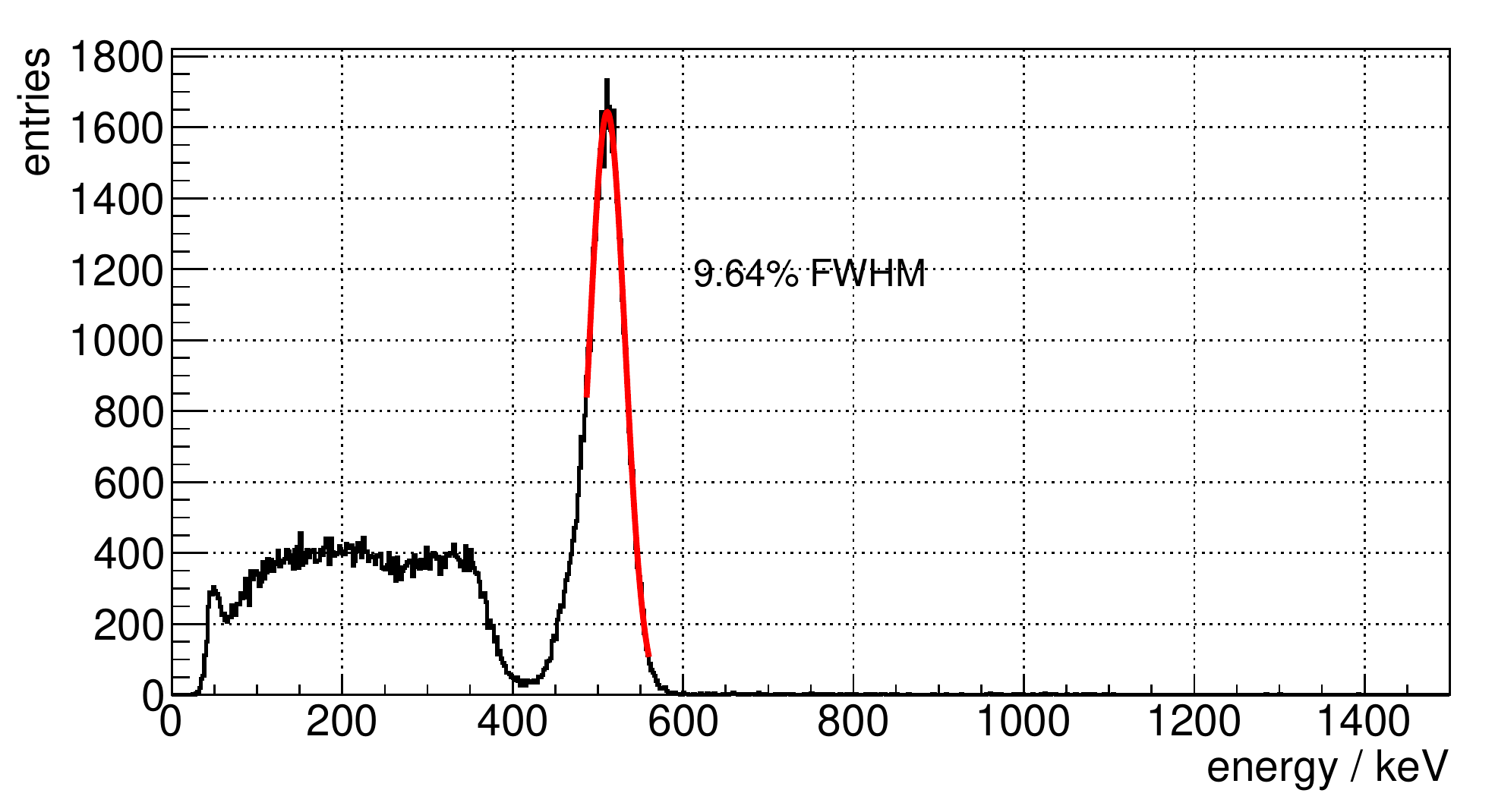}{width=1.00\columnwidth}{fig:coincident_energy_spectrum}
  \caption{\protect\subref{fig:crystal_0_energy_value_spectrum} Uncorrected energy value spectrum for one of the two crystals.
    \protect\subref{fig:crystal_0_energy_spectrum} The saturation-corrected energy spectrum for the same crystal.
    \protect\subref{fig:coincident_energy_spectrum} Energy spectrum of coincident events.
     The measurements were conducted at an overvoltage of $V_\text{OV} = \SI{3.25}{\volt}$ and a discriminator threshold of $\mathsf{vth\_t1} = 20$.
  }
  \label{fig:energyspectra}
\end{figure}

\begin{figure}[tb]
    \centering
    \tikzsubfig{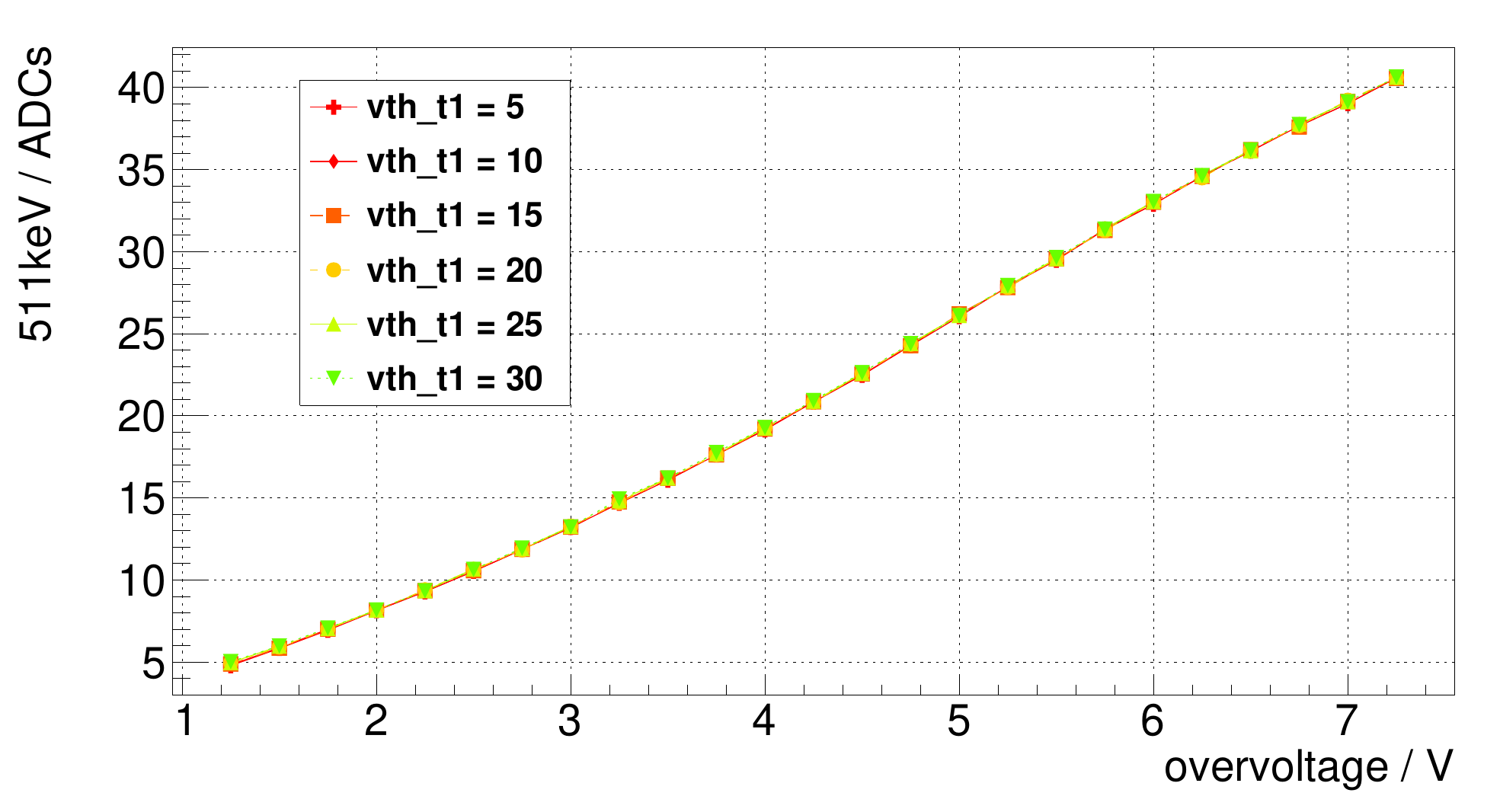}{width=1.00\columnwidth}{fig:511keVpeak_vs_bias}
    \\
    \tikzsubfig{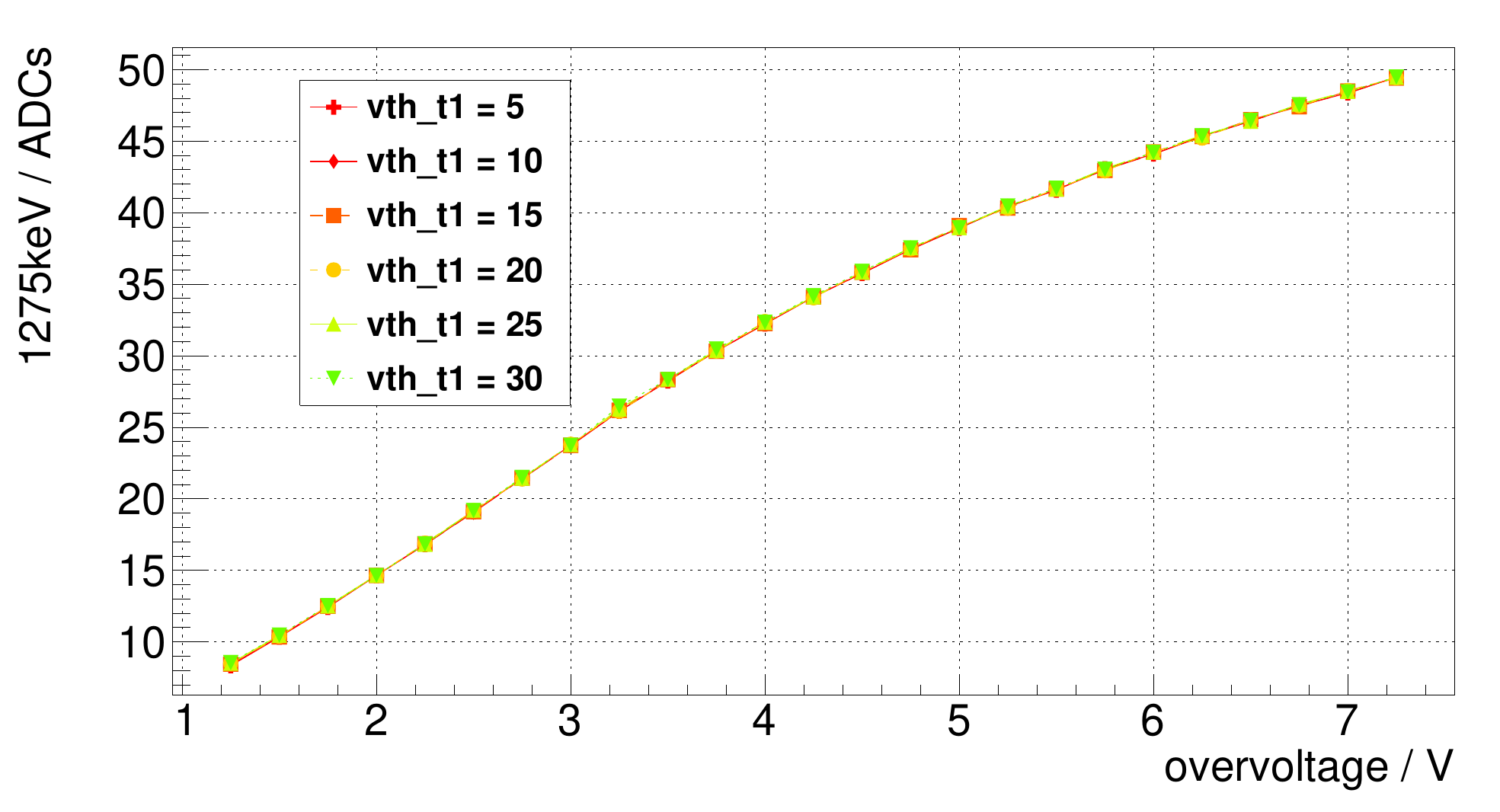}{width=1.00\columnwidth}{fig:1275keVpeak_vs_bias}
    \caption{Energy value of \protect\subref{fig:511keVpeak_vs_bias} the \SI{511}{\keV} peak and the \protect\subref{fig:1275keVpeak_vs_bias} the \SI{1275}{\keV} over overvoltage for one of the two crystals for different values of the discriminator threshold $V_\mathsf{th\_T1}$.}
    \label{fig:511keVpeak_and_1275keVpea_vs_bias}
\end{figure}

The locations of the \SI{511}{\keV} and \SI{1275}{\keV} photopeaks (\autoref{fig:crystal_0_energy_value_spectrum}) showed the expected monotonous increase with a rising overvoltage and are independent of the used discriminator threshold  $V_\mathsf{th\_T1}$.
While the \SI{511}{\keV} peak position showed an almost linear increase with rising overvoltage, the \SI{1275}{\keV} peak position showed a stronger saturation effect (\autoref{fig:511keVpeak_and_1275keVpea_vs_bias}).
Only minor differences in the peak positions were observed for the two detectors.

\begin{figure}[tb]
	\centering
	\includegraphics[width=1.00\columnwidth]{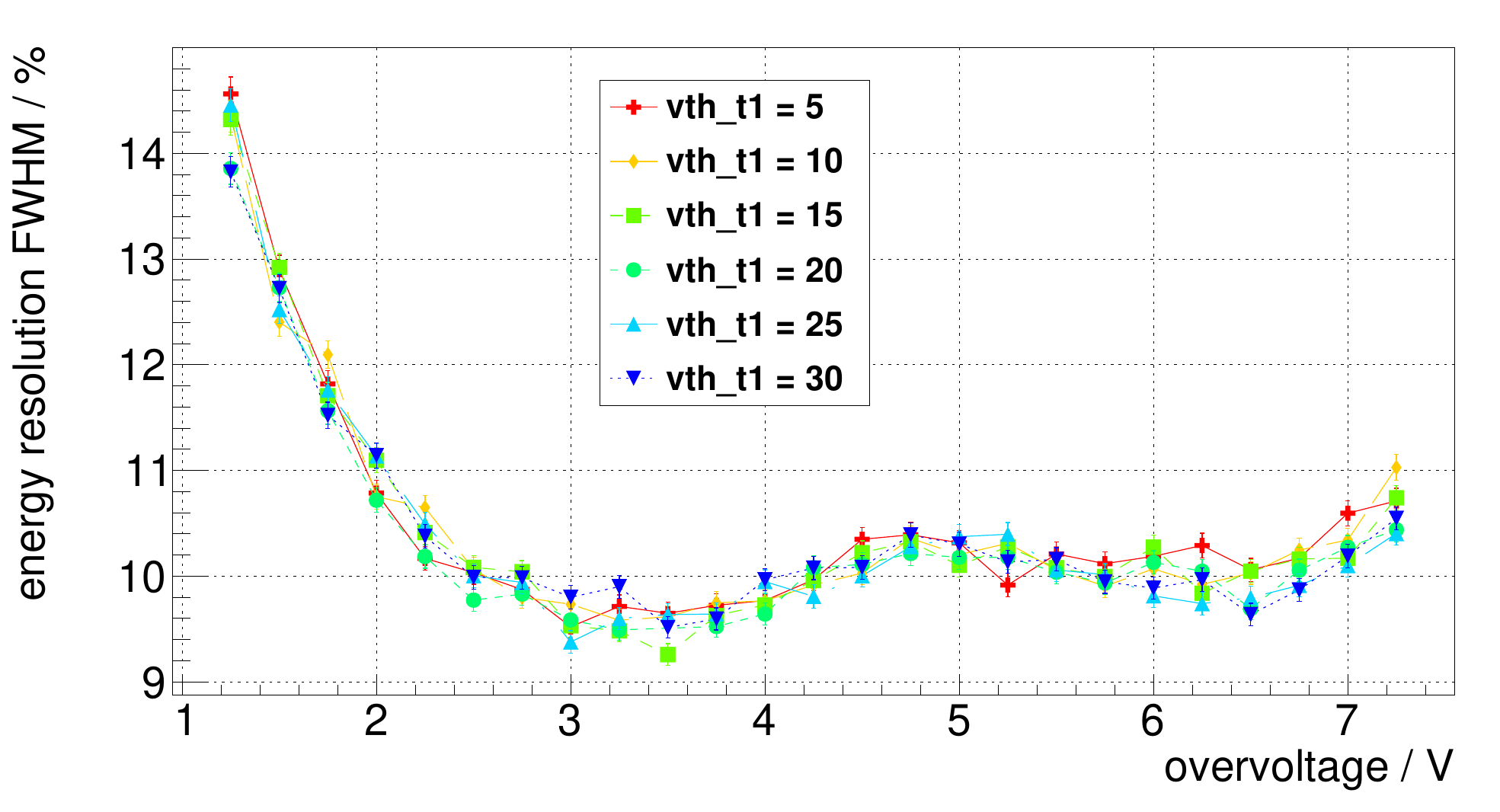}
	\caption{Energy resolution of the \SI{511}{\keV} peak over overvoltage for the measured values of the discriminator threshold $V_\mathsf{th\_T1}$ at an ambient temperature of \SI{16}{\celsius}.}
	\label{fig:ERes_vs_bias}
\end{figure}

The energy resolution for coincident events (\autoref{fig:coincident_energy_spectrum}) for the \SI{511}{\keV} photopeak improved from the lowest values of $V_\text{OV}$ up to $V_\text{OV} = \SI{3.50}{\volt}$.
At this voltage, the energy resolution was determined to be \SI{9.7}{\percent}~FWHM and slightly deteriorated for higher voltages to values of about \SIrange{9.8}{10.2}{\percent}~FWHM and worsened for $V_\text{OV} > \SI{6.50}{\volt}$ more quickly to \SI{11}{\percent}~FWHM.
No significant difference was observed for different values of the discriminator threshold $V_\mathsf{th\_T1}$ (\autoref{fig:ERes_vs_bias}).

\begin{figure}[tb]
  \centering
  \tikzsubfig{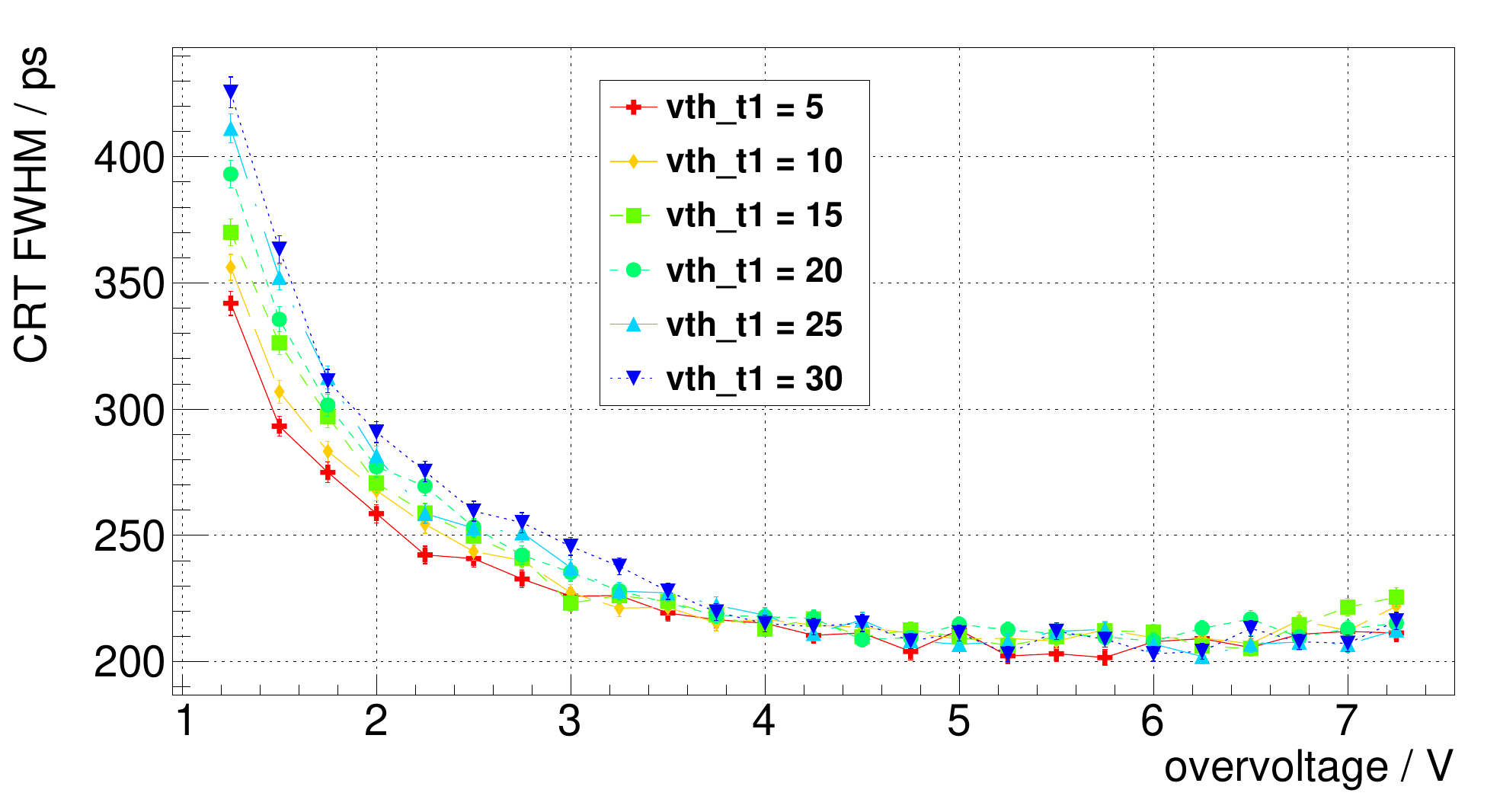}{width=1.00\columnwidth}{fig:CRT-FWHM_vs_bias}
  \\
  \tikzsubfig{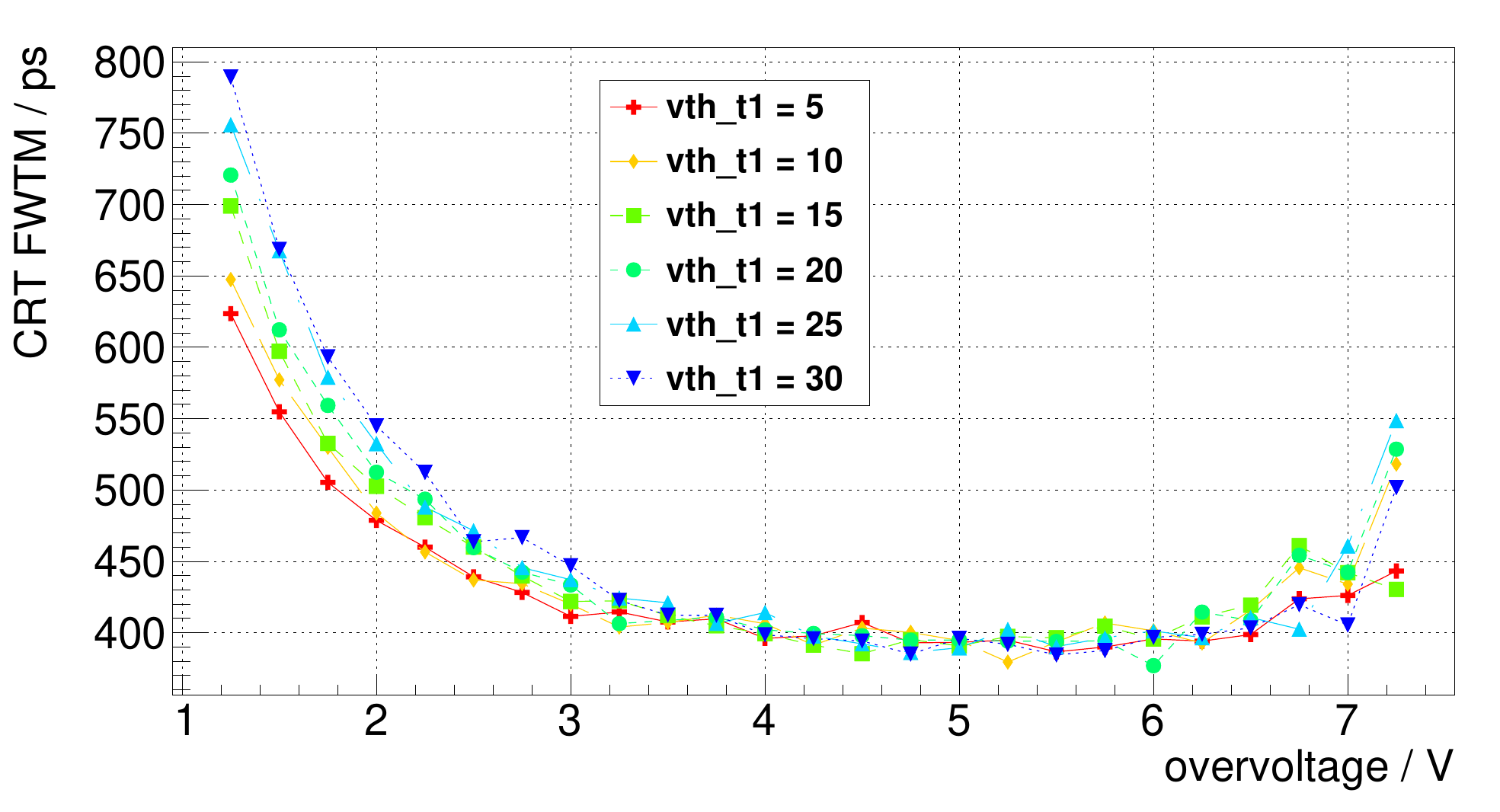}{width=1.00\columnwidth}{fig:CRT-FWTM_vs_bias}
  \caption{\protect\subref{fig:CRT-FWHM_vs_bias} FWHM and \protect\subref{fig:CRT-FWTM_vs_bias} FWTM of the time difference histogram over overvoltage for the measured values of the discriminator threshold $V_\mathsf{th\_T1}$ at an ambient temperature of \SI{16}{\celsius}.}
  \label{fig:CRT-FWHM_and_FWTM_vs_bias}
\end{figure}

The CRT showed a clear improvement with increasing voltage in the range of $V_\text{OV} = \SIrange{1.25}{5.00}{\volt}$.
The optimum was reached for about $V_\text{OV} = \SIrange{5.00}{6.00}{\volt}$ with approximate values of \SI{210}{\pico\second}~FWHM and \SI{400}{\pico\second}~FWTM.
For lower voltages, a clear difference was measured for different settings of the $\mathsf{D\_T1}$ discriminator with lower values of $V_\mathsf{th\_T1}$ showing a better CRT performance.
For voltages above approximately $V_\text{OV} > \SI{4.25}{\volt}$, no difference could be measured.
If the voltage was increased above  $V_\text{OV} > \SI{6.00}{\volt}$, the CRT showed a degradation
(\autoref{fig:CRT-FWHM_and_FWTM_vs_bias}).

\begin{figure}[tb]
	\centering
	\includegraphics[width=1.00\columnwidth]{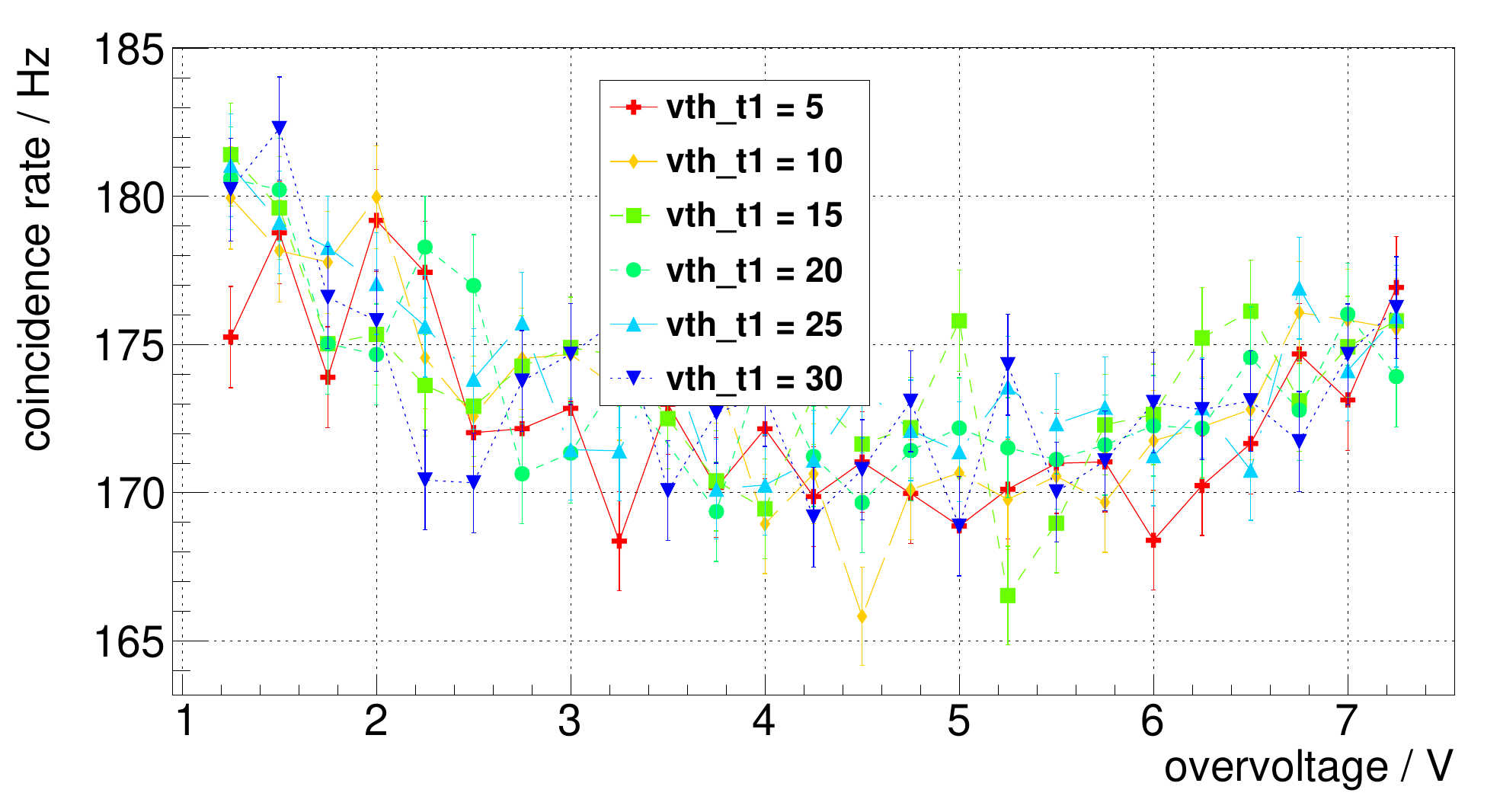}
	\caption{Coincidence rate over overvoltage for the measured values of the discriminator threshold $V_\mathsf{th\_T1}$ at an ambient temperature of \SI{16}{\celsius}.}
	\label{fig:Coincs_vs_bias}
\end{figure}

The coincidence rate showed only small variations of about \SI{5}{\percent} in the evaluated parameter range.
At an overvoltage of  $V_\text{OV} = \SIrange{5.00}{6.00}{\volt}$, a coincidence rate of  approximately \SI{172}{\text{cps}} was measured.
The different settings of the $\mathsf{D\_T1}$ discriminator did not show a significant influence on the measured coincidence rate (\autoref{fig:Coincs_vs_bias}).

\begin{figure}[tb]
  \centering
  \tikzsubfig{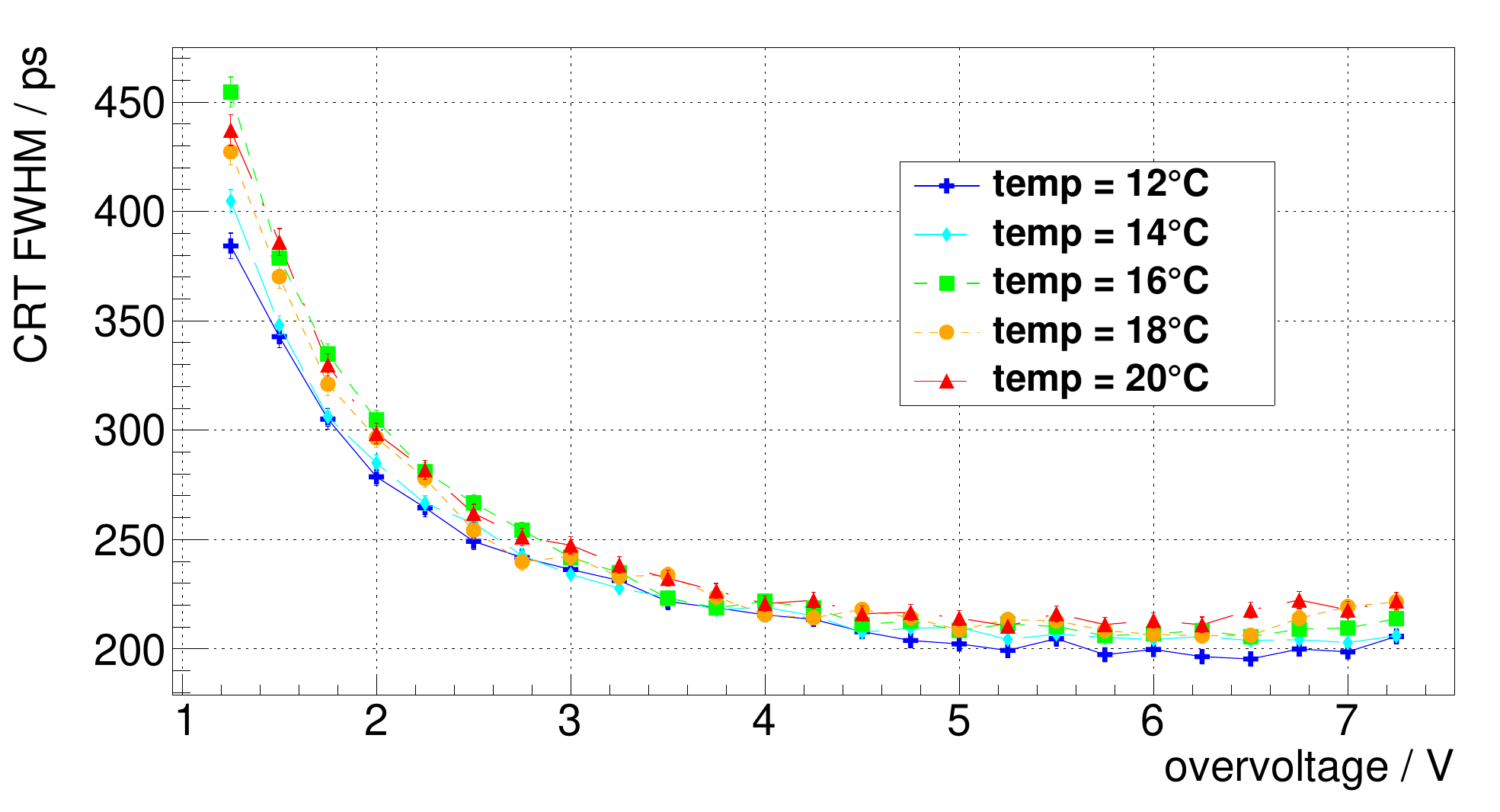}{width=1.00\columnwidth}{fig:temperature_dependency_t130_CRT_FWHM}
  \\
  \tikzsubfig{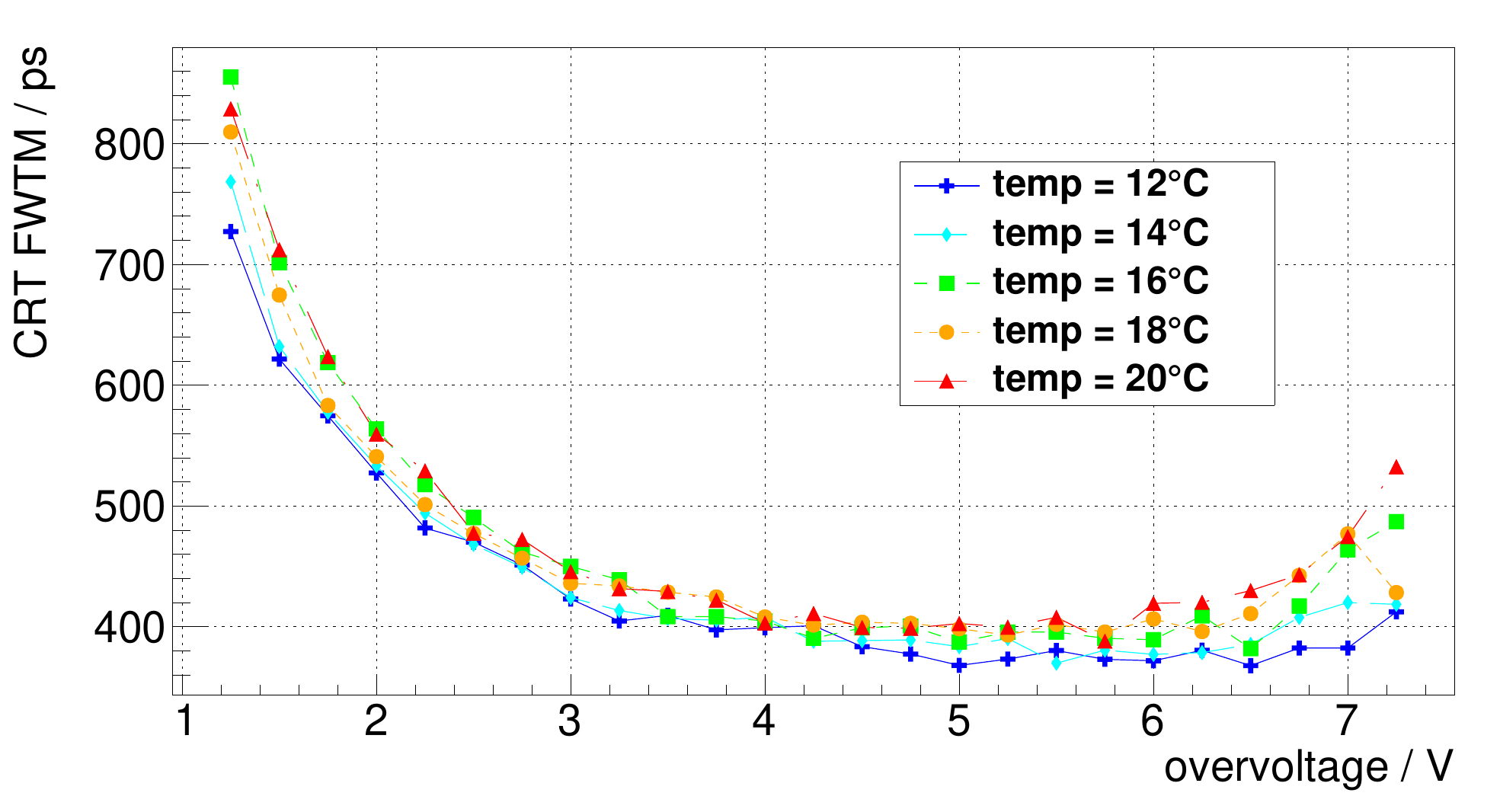}{width=1.00\columnwidth}{fig:temperature_dependency_t130_CRT_FTHM}
  \caption{\protect\subref{fig:temperature_dependency_t130_CRT_FWHM} FWHM and \protect\subref{fig:temperature_dependency_t130_CRT_FTHM} FWTM of the time difference histogram plotted over the applied overvoltage for the different ambient temperatures investigated at a discriminator threshold of $\mathsf{vth\_t1} = 30$.}
  \label{fig:temperature_dependency_t130_CRT_FWHM_and_FWTM}
\end{figure}

\begin{figure}[tb]
	\centering
	\includegraphics[width=1.00\columnwidth]{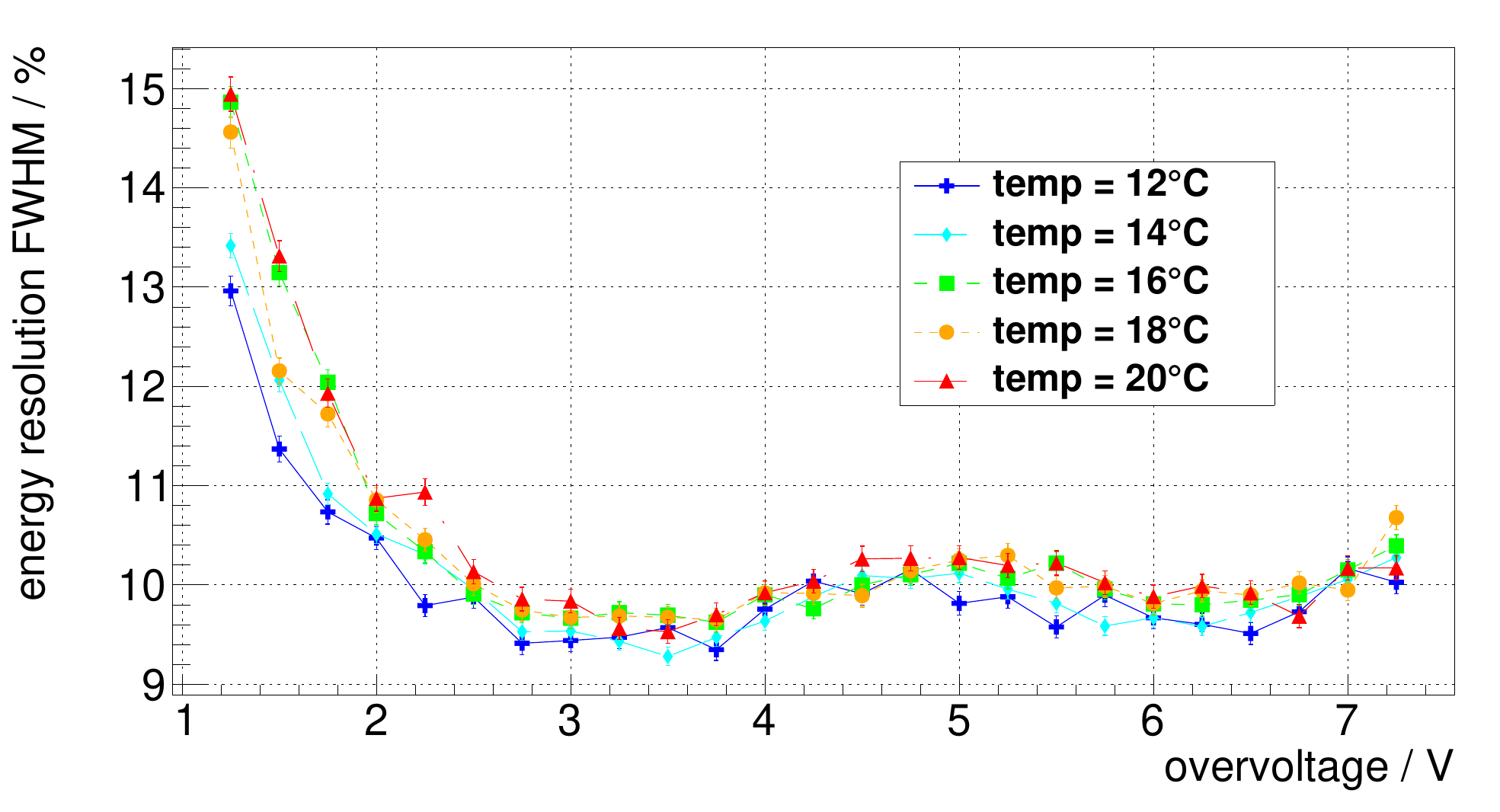}
	\caption{Temperature dependency of the energy resolution of the \SI{511}{\keV} peak plotted over the applied overvoltage at a discriminator threshold of $\mathsf{vth\_t1} = 30$.}
	\label{fig:temperature_dependency_t130_eRes}
\end{figure}

The CRT and the energy resolution showed a performance improvement for lower temperatures and were observed to be stable up to higher voltages compared to the default temperature of \SI{16}{\celsius}.
At \SI{12}{\celsius}, a CRT of \SI{195}{\ps}~FWHM (\SI{370}{\pico\second}~FWTM) and an energy resolution of about \SI{9.5}{\percent}~FWHM were measured in the voltage range of $V_\text{OV} = \SI{6.50}{\volt}$.
For this temperature, no energy resolution degradation was observed towards the highest values of $V_\text{OV}$.
Higher temperatures showed worse performance values over the whole range of voltages measured and an earlier degradation of the energy resolution and CRT, especially the FWTM, towards the highest voltages applied
(\autoref{fig:temperature_dependency_t130_CRT_FWHM_and_FWTM} and \autoref{fig:temperature_dependency_t130_eRes}).

\begin{figure}[tb]
  \centering
  \tikzsubfig{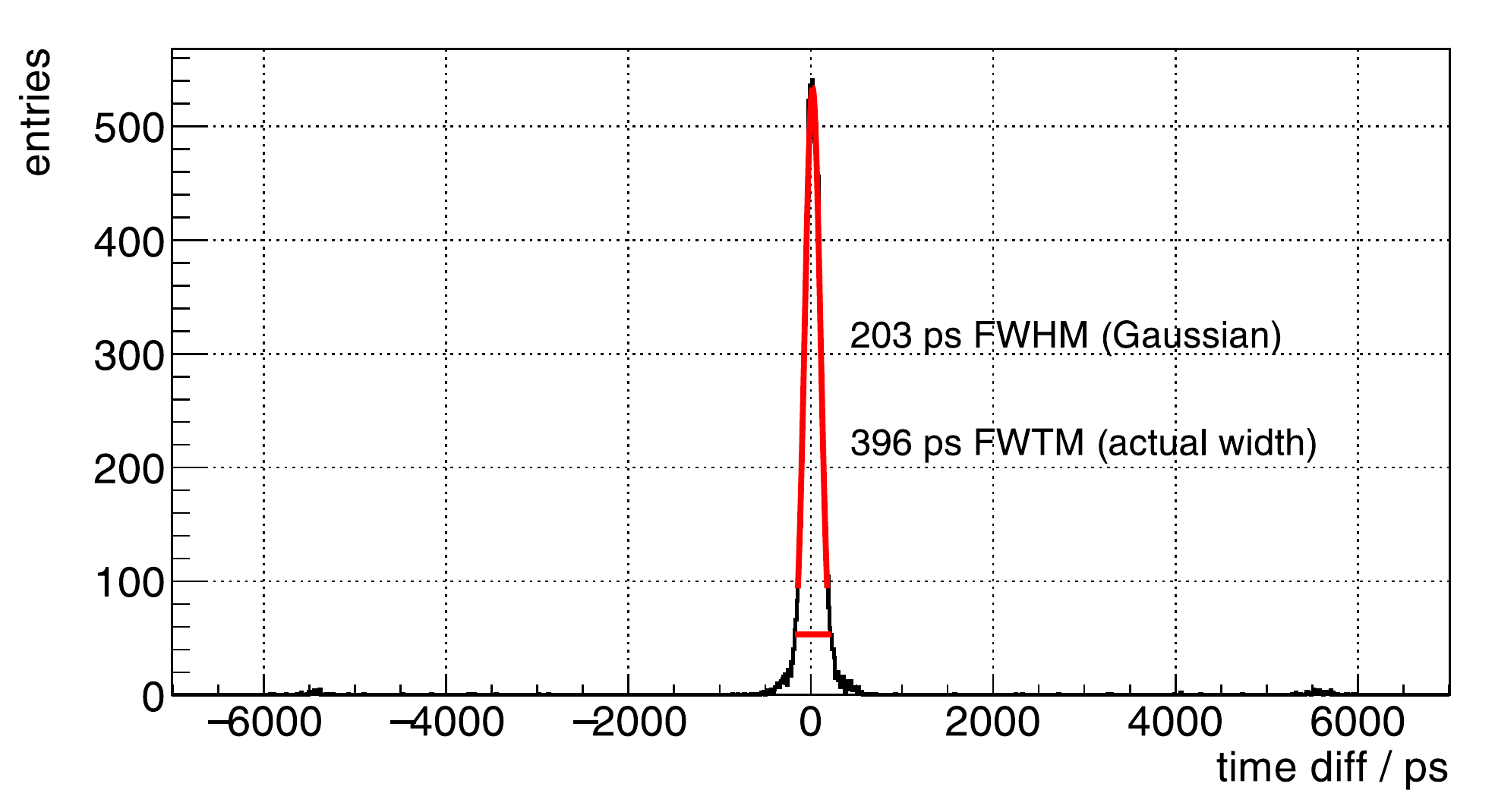}{width=1.00\columnwidth}{fig:time_diff_spectrum_OV6V_t130}
  \\
  \tikzsubfig{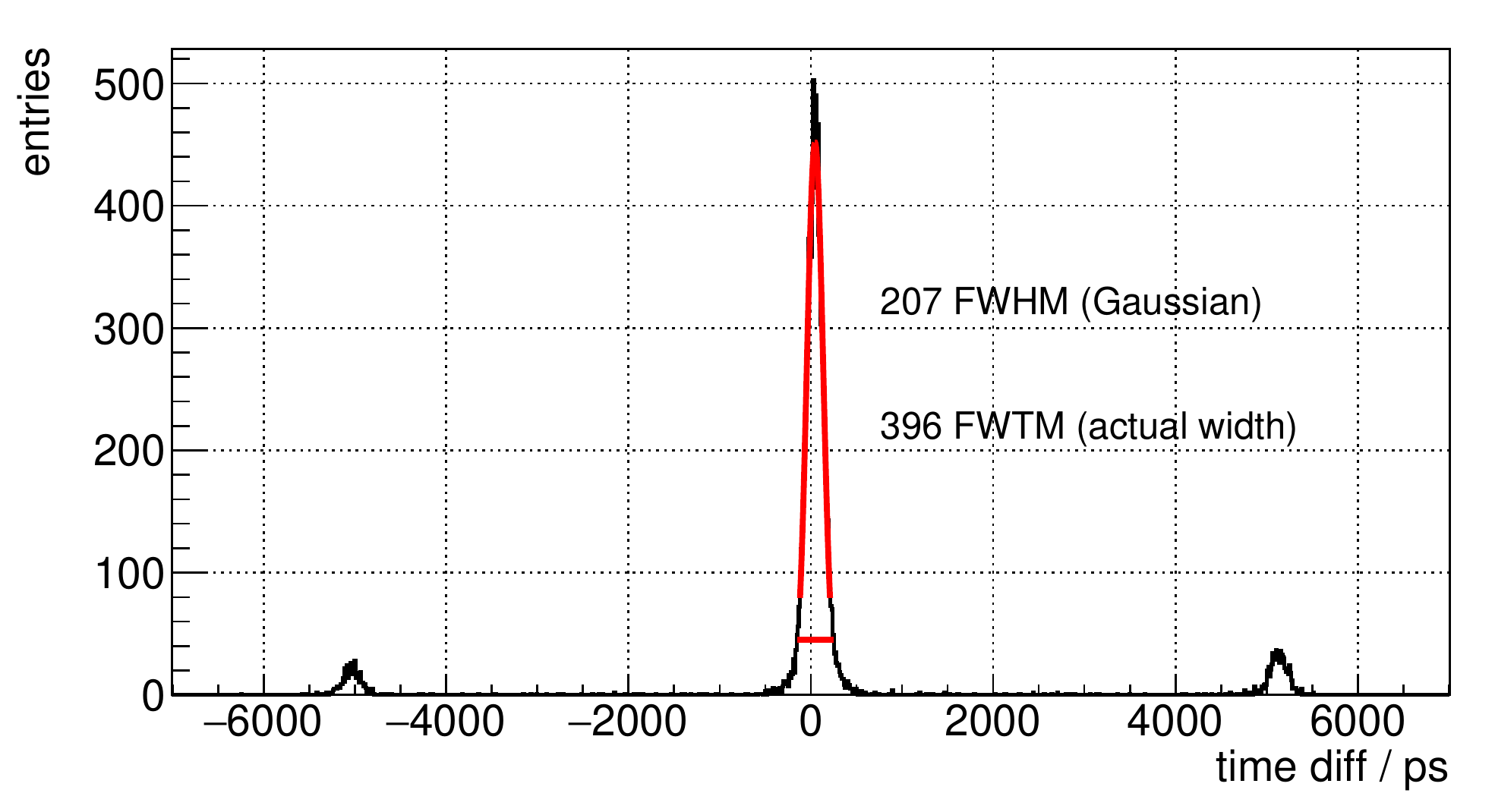}{width=1.00\columnwidth}{fig:time_diff_spectrum_OV6V_t15}
  \caption{Time difference spectrum for an overvoltage $V_\text{OV} = \SI{6}{\volt}$ and discriminator threshold of \protect\subref{fig:time_diff_spectrum_OV6V_t130} $\mathsf{vth\_t1} = 30$ and \protect\subref{fig:time_diff_spectrum_OV6V_t15} $\mathsf{vth\_t1} = 5$, both for $\mathsf{fe\_delay}$ set to \num{14} which should correspond to $t_\mathsf{delay\_T1} = \protect\SI{5.80}{\nano\second}$.}
  \label{fig:time_diff_spectrum_OV6V_t130_and_t15}
\end{figure}

\begin{figure}[tb]
    \centering
    \includegraphics[width=1.00\columnwidth]{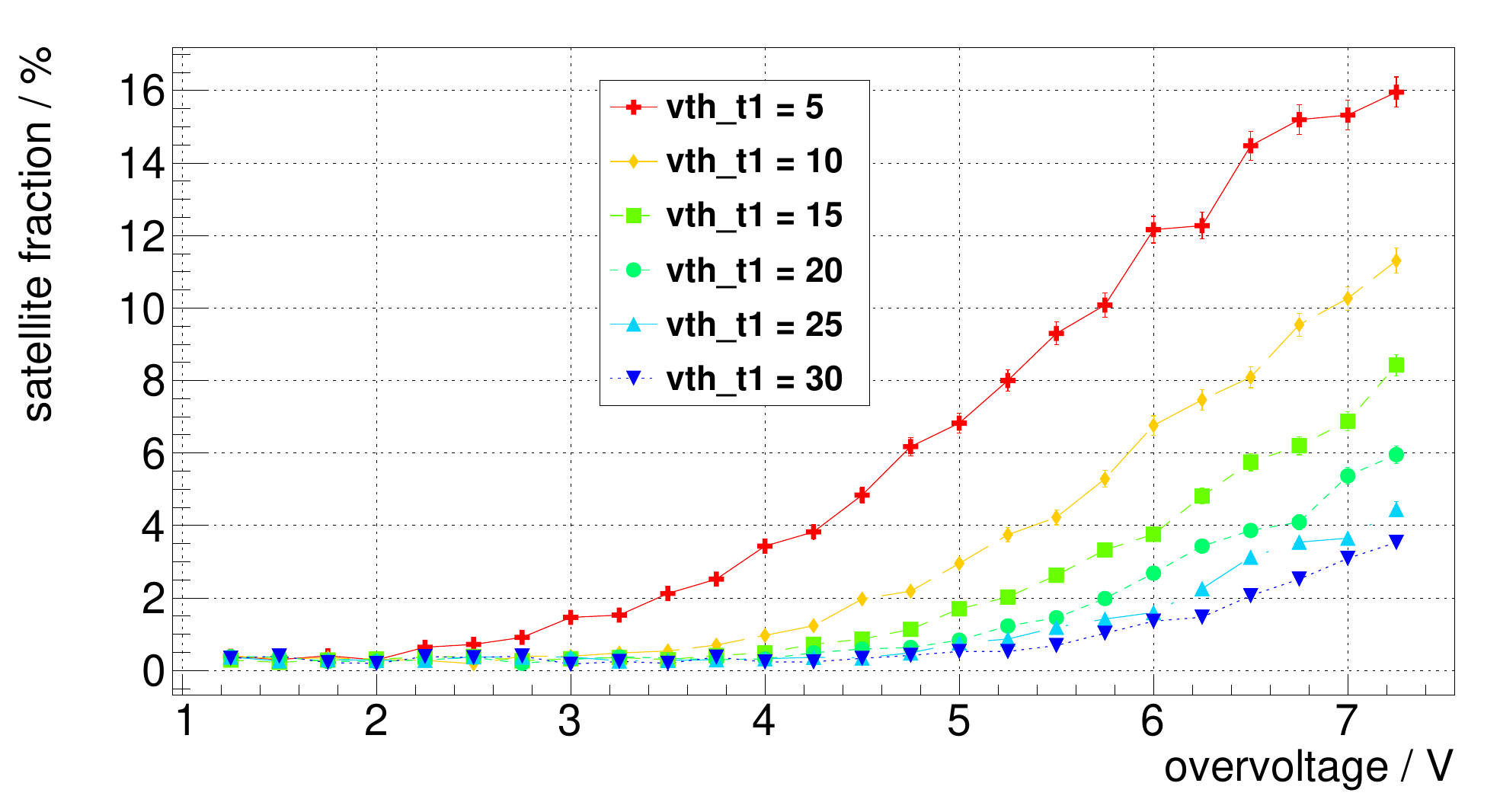}
    \caption{Satellite peak fraction over overvoltage for the measured values of the discriminator threshold $V_\mathsf{th\_T1}$ at an ambient temperature of \SI{16}{\celsius}.}
    \label{fig:satellitefraction_over_ov}
\end{figure}

\begin{figure}[tb]
  \centering
  \includegraphics[width=1.00\columnwidth]{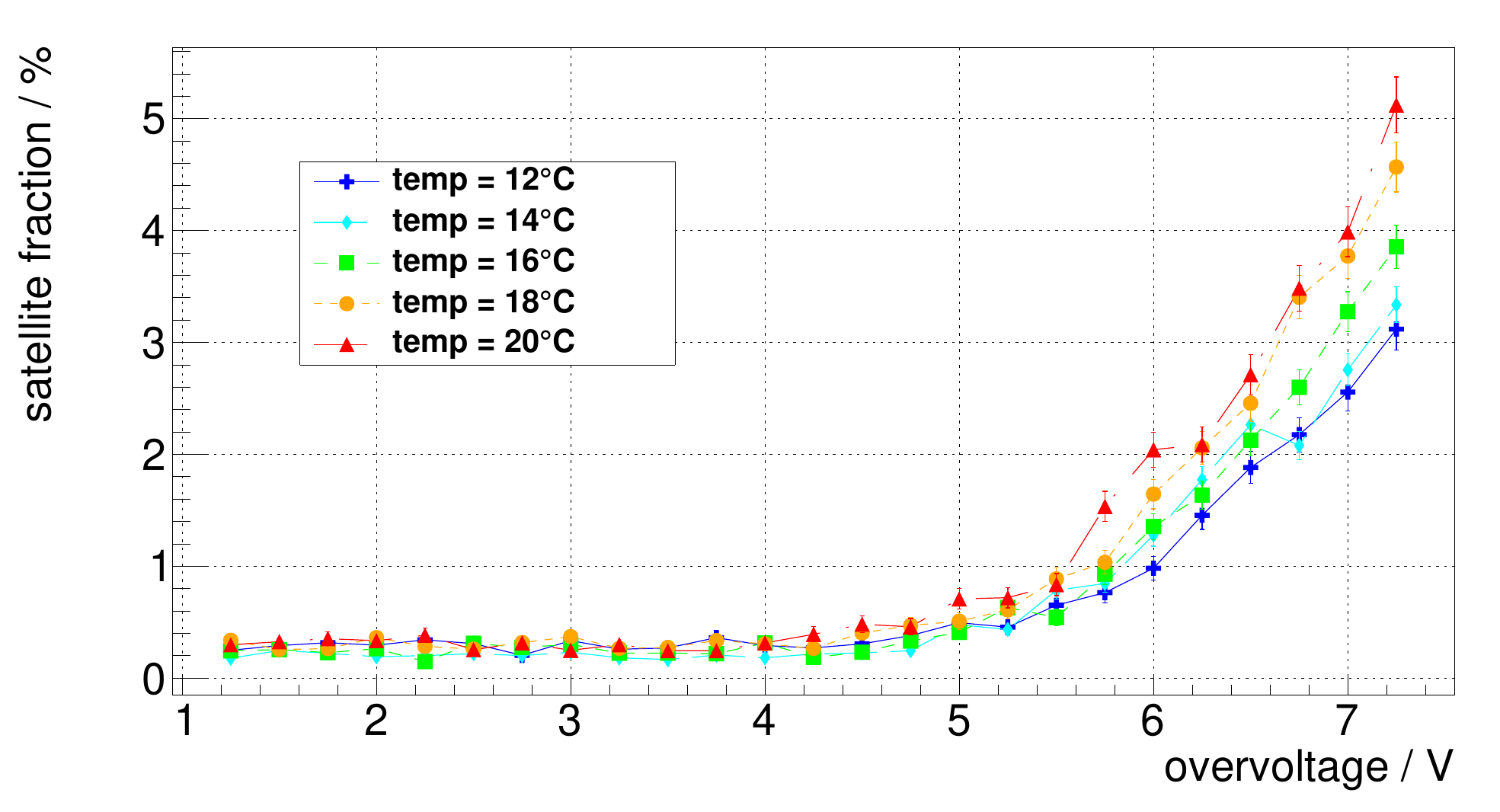}
  \caption{Temperature dependency of the satellite peak fraction over overvoltage for the different ambient temperatures investigated at a discriminator threshold of $\mathsf{vth\_t1} = 30$.}
  \label{fig:satellitefraction_over_ov_temperatures}
\end{figure}

\begin{figure*}[tb]
    \centering
    \tikzsubfig{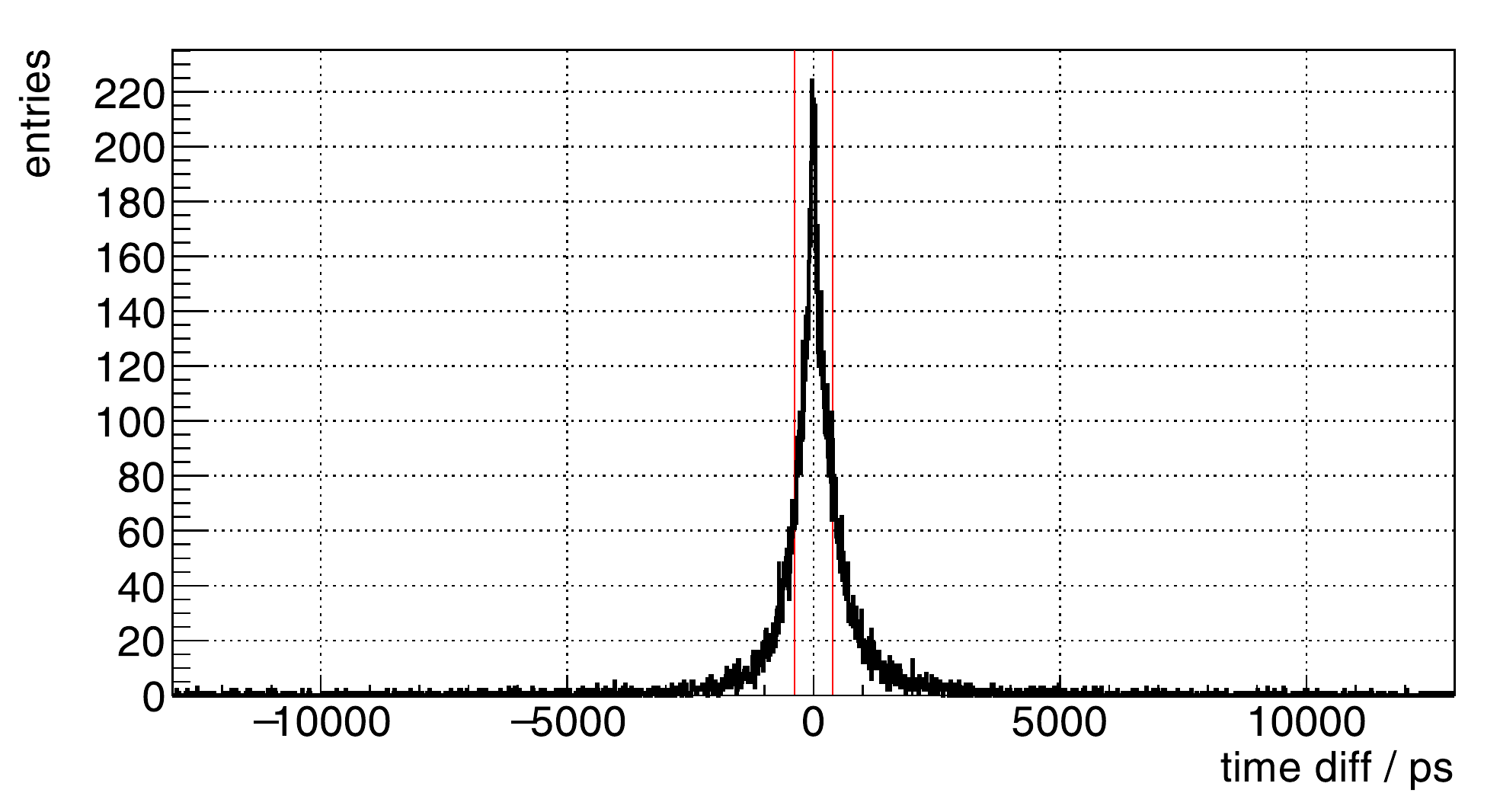}{width=0.50\textwidth}{subfig:OV10_distance10mm_delay12}
    \tikzsubfig{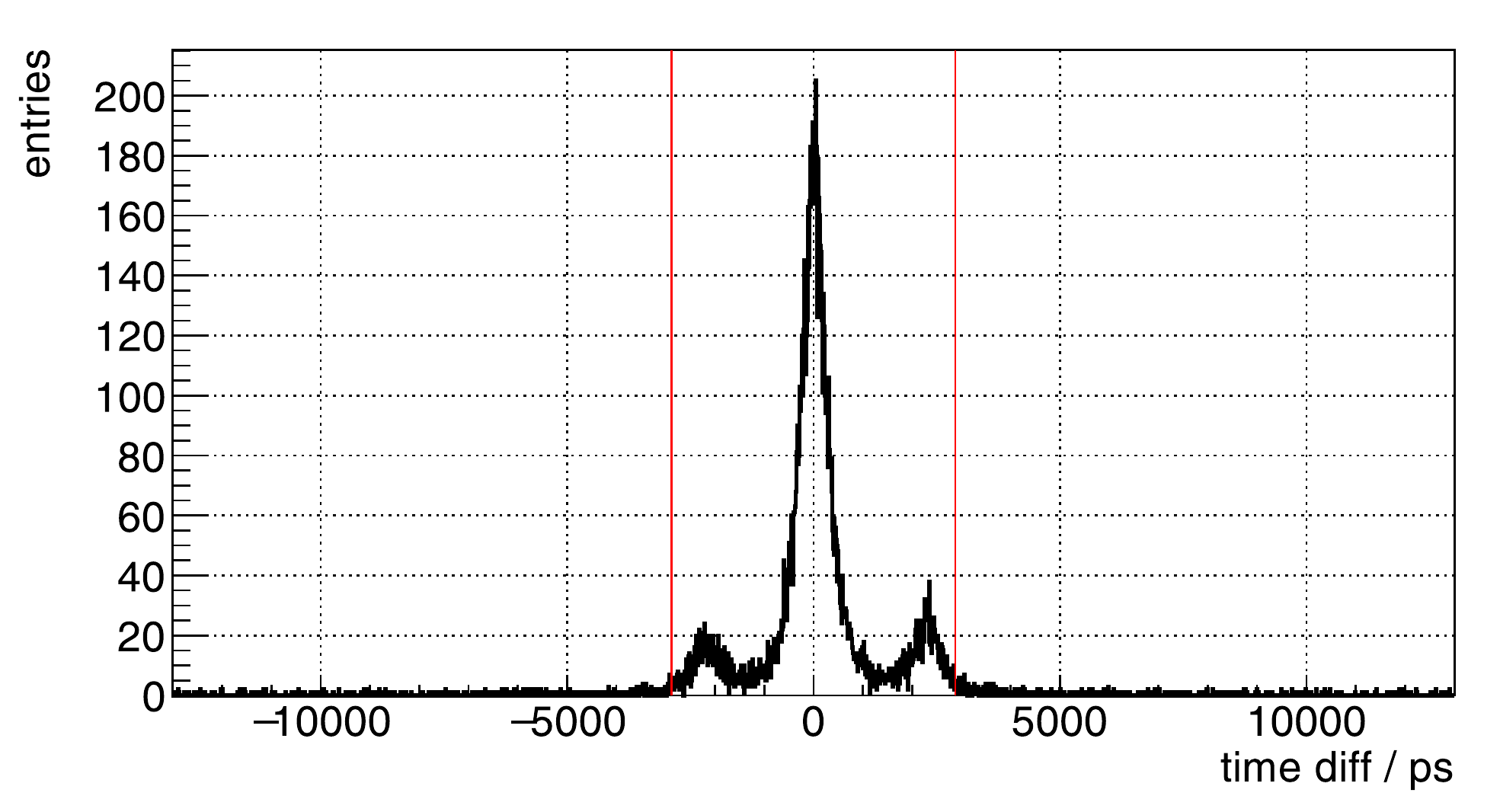}{width=0.50\textwidth}{subfig:OV10_distance10mm_delay13}
    \\
    \tikzsubfig{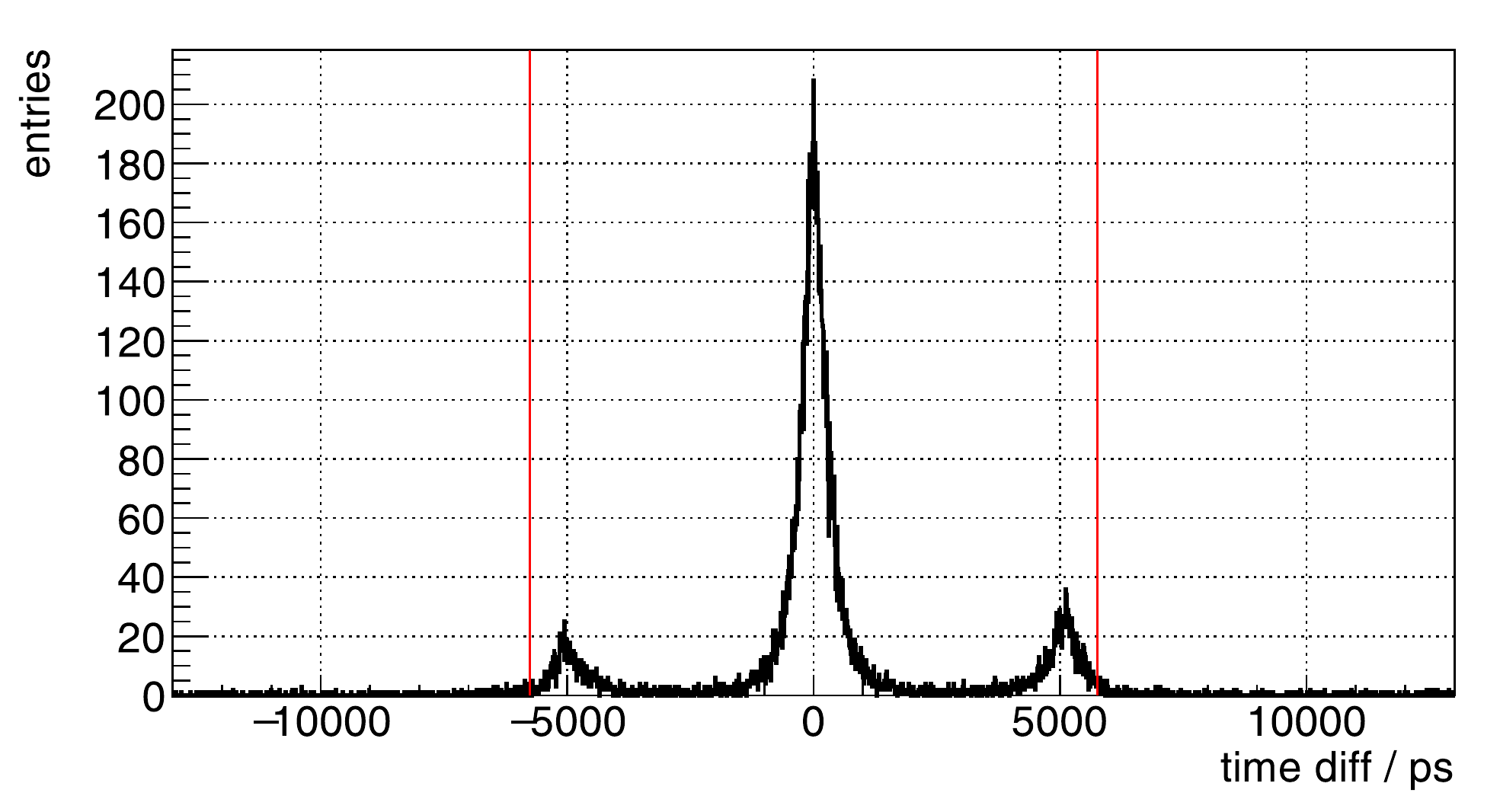}{width=0.50\textwidth}{subfig:OV10_distance10mm_delay14}
    \tikzsubfig{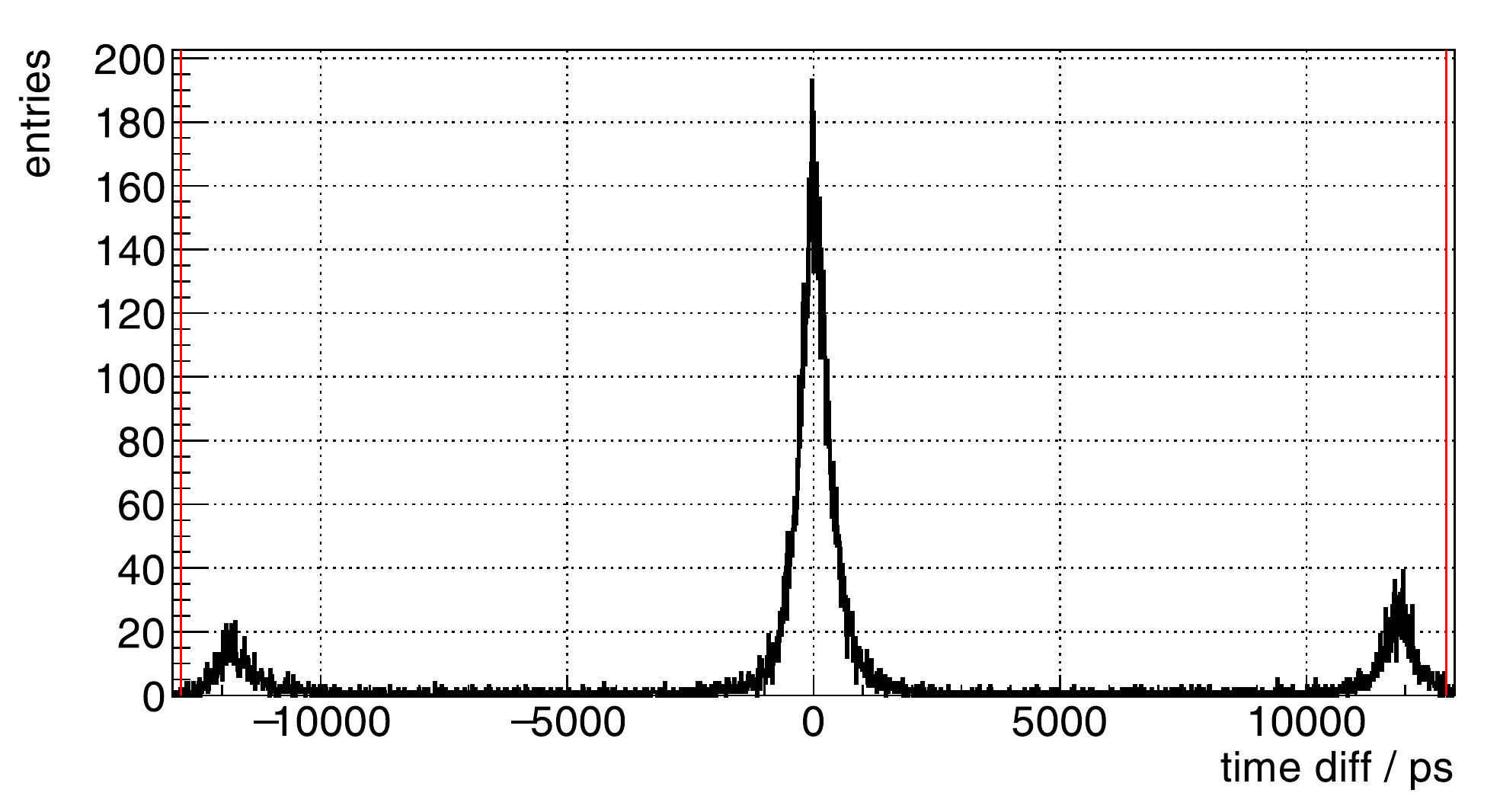}{width=0.50\textwidth}{subfig:OV10_distance10mm_delay11}
    \caption{Time difference spectra for different settings of the $\mathsf{delay\_T1}$ element. The theoretical value provided by PETsys is marked with red lines.
        \protect\subref{subfig:OV10_distance10mm_delay12} $\mathsf{fe\_delay}$ set to \num{12} which should correspond to $t_\mathsf{delay\_T1} = \protect\SI{0.39}{\nano\second}$.
        \protect\subref{subfig:OV10_distance10mm_delay13} $\mathsf{fe\_delay}$ set to \num{13} which should correspond to $t_\mathsf{delay\_T1} = \protect\SI{2.95}{\nano\second}$.
        \protect\subref{subfig:OV10_distance10mm_delay14} $\mathsf{fe\_delay}$ set to \num{14} which should correspond to $t_\mathsf{delay\_T1} = \protect\SI{5.80}{\nano\second}$.
        \protect\subref{subfig:OV10_distance10mm_delay11} $\mathsf{fe\_delay}$ set to \num{11} which should correspond to $t_\mathsf{delay\_T1} = \protect\SI{12.90}{\nano\second}$.
    }
    \label{fig:delay_scan}
\end{figure*}

We observed satellite peaks in many of the time difference spectra.
The effect was more prominent for higher overvoltages and lower values of $V_\mathsf{th\_T1}$ (\autoref{fig:time_diff_spectrum_OV6V_t130} and \autoref{fig:time_diff_spectrum_OV6V_t15}).
For a quantative evaluation of all measurements conducted with the default settings, we defined coincidences with a time difference larger than \SI{2.5}{\ns} to contribute to the satellite fraction.
A baseline of the satellite fraction of about \SI{0.3}{\percent} was observed.
All applied values for $\mathsf{vth\_t1}$ showed a significant increase of the satellite fraction above the baseline when surpassing a threshold-specific overvoltage.
For $\mathsf{vth\_t1} = 5$, the satellite fraction increased above the baseline for overvoltages higher than \SI{2.0}{\volt} and reached the highest value of about \SI{16}{\percent} for the maximum overvoltage (\autoref{fig:satellitefraction_over_ov}).
The satellite peak fraction increased for higher temperatures when all other parameters where kept constant (\autoref{fig:satellitefraction_over_ov_temperatures}).

The location of the satellite peaks was observed to be correlated with the programmed length of the $\mathsf{delay\_T1}$ element (\autoref{fig:delay_scan}).
For the shortest theoretical value of $t_\mathsf{delay\_T1} = \SI{0.39}{ns}$ the satellite peaks were presumably merged into the main peak and could not be resolved anymore (\autoref{subfig:OV10_distance10mm_delay12}).

\begin{figure}[tb]
	\centering
	\includegraphics[width=1.00\columnwidth]{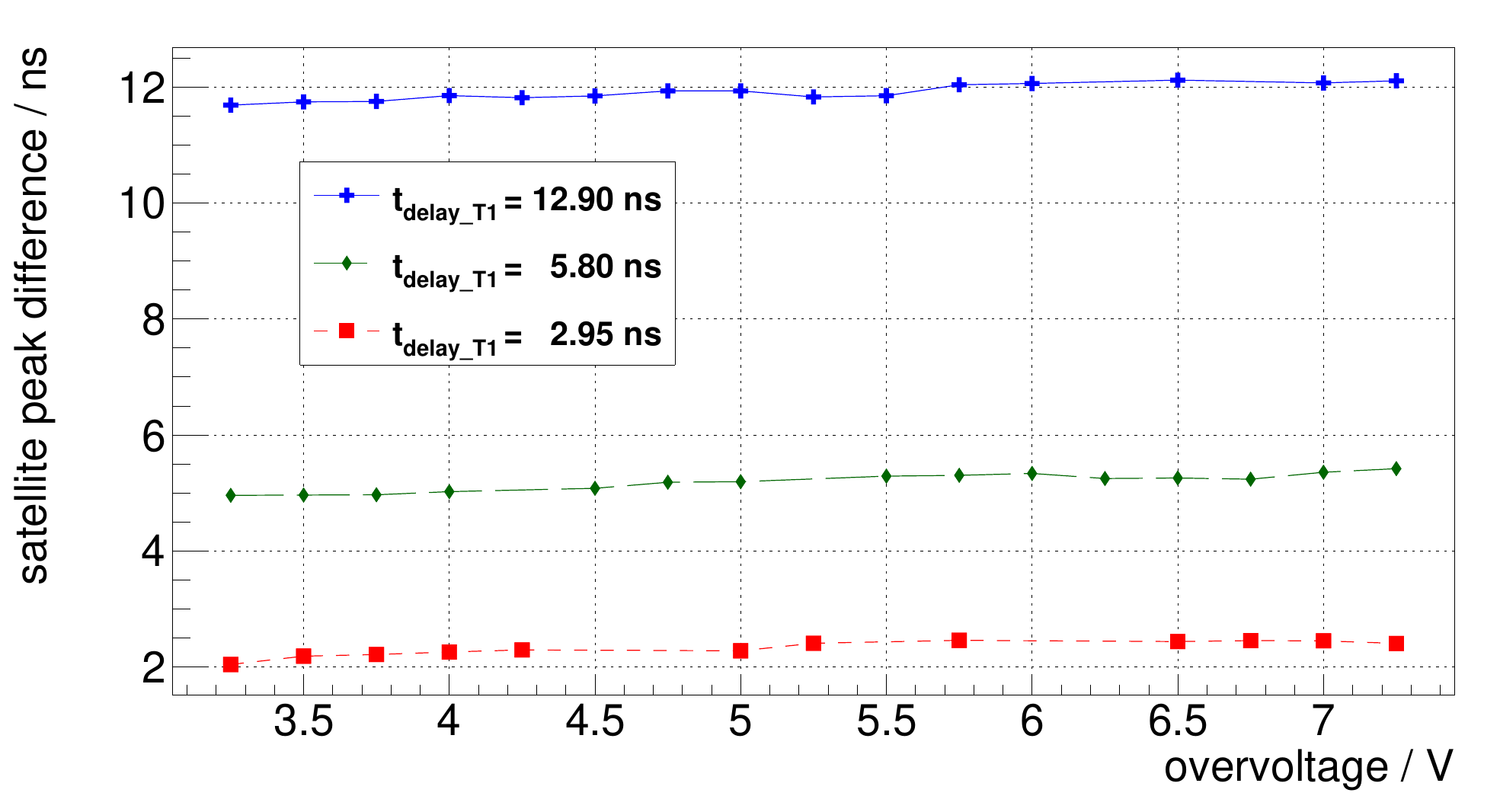}
	\caption{Peak position difference for different overvoltage and delay settings and a discriminator threshold of $\mathsf{vth\_t1} = 1$.}
	\label{fig:real_shift_vs_ov}
\end{figure}

\autoref{fig:real_shift_vs_ov} shows the difference between the main peak and the satellite peak locations as a function of overvoltage for the three higher values of $t_\mathsf{delay\_T1}$ that could still be resolved.
We observed an increase in the satellite peak distance to the main peak in the time difference spectrum with increasing voltage.

\section{Discussion and Conclusions}

\begin{table*}[t]
	\caption{Combined SiPM ASIC performance obtained by comparable setups.}
	
	\centering
	\begin{tabularx}{\textwidth}{llll@{}rrrrr}
		\hline
		ASIC & SiPM & \multicolumn{3}{l}{scintillator} & temp. / \si{\celsius} & CRT~FWHM / \si{\ps} & e. res.~FWHM~ / \si{\percent} & ref. \\
		\hline
		TOFPET2 & KETEK PM3325-WB-A0 		& LYSO & $\SI{3.0x3.0}{\mm} \times$ &  \SI{5}{\mm} & 18 &  229 & 10.5 & \cite{TOFPET2EvaluationReport,bugalho2017ExperimentalResultsTOFPET2} \\
		TOFPET2 & KETEK PM3325-WB-A0 		& LYSO & $\SI{3.0x3.0}{\mm} \times$ &  \SI{5}{\mm} & 18 &  202 & 8.0 &  \cite{lamprou2018CharacterizationTOFPET} \\
		TOFPET1 & FBK NUV-HD 						& LYSO & $\SI{3.1x3.1}{\mm}\times$ & \SI{15}{\mm} & n.a. &  291 & 12.5 &  \cite{niknejad2016TOFPETvalidation} \\
		STiC v3 & S12643-050CN(X) MPPC 	& LYSO & $\SI{3.1x3.1}{\mm}\times$ & \SI{15}{\mm} & 22 &  214 & n.a. & \cite{chen2014DedicatedReadoutASIC,stankova2015STIC3} \\
		ToT ASIC & S13361-3050AE-08 MPPC 		& GFAG & $\SI{2.0x2.0}{\mm}\times$ &  \SI{2}{\mm} & 25 &  200 & 13.5 & \cite{orita2017CurrentToTASIC} \\
		PETA3 & FBK 									  & LYSO & $\SI{3.0x3.0}{\mm} \times$ &  \SI{5}{\mm} & 20 &  190 & 12.0 & \cite{piemonte2012PerformancePETA3ASIC,sacco2013PETA4} \\
		PETA3 & FBK 									  & LYSO & $\SI{3.0x3.0}{\mm} \times$ & \SI{15}{\mm} & 20 &  250 & n.a. & \cite{piemonte2012PerformancePETA3ASIC} \\
		PETA5 & FBK RGB-HD 							& LYSO & $\SI{2.5x2.5}{\mm} \times$ & \SI{10}{\mm} & 20 &  205 & n.a. & \cite{sacco2015PETA5} \\
		Petiroc & KETEK  								& LYSO & $\SI{3.0x3.0}{\mm} \times$ &  \SI{3}{\mm} & n.a. &  n.a. & 9.5 & \cite{fleury2013Petiroc} \\
		Triroc & ADVANSID NUV  				  & LYSO & $\SI{3.0x3.0}{\mm} \times$ &  \SI{8}{\mm} & 23 &  420 & 10.7 & \cite{sportelli2016TRIMAGEInitialResults} \\
		Triroc & ADVANSID NUV  					& LYSO & $\SI{3.0x3.0}{\mm} \times$ & \SI{10}{\mm} & n.a. &  433 & 11.0 & \cite{ahmad2016Triroc} \\
		\hline
	\end{tabularx}
	\label{tab:comparablesetups}
\end{table*}

We expect only minor influence of the energy linearization assuming a baseline at zero and only statistical saturation effects of the SiPM on the evaluated energy resolution.
The compton edges can be identified in the linearized energy spectrum in agreement with the theoretical positions (\autoref{fig:energyspectra}).
Remaining non-linearities in the range of the photopeak are expected to be minor and have only a small influence on the peak width.

The results obtained with small LYSO crystals, read out by a KETEK PM3325-WB-A0 SiPM each coupled to a single input channel of a TOFPET2 ASIC, obtained in this work are in agreement with the previously published results by PETsys \cite{TOFPET2EvaluationReport,bugalho2017ExperimentalResultsTOFPET2}. They demonstrate the state-of-the-art performance of the ASIC under these artificial test conditions.
Other groups using the same setup report results in the same order of magnitude \cite{lamprou2018CharacterizationTOFPET}.
The results obtained with TOFPET2 ASIC are comparable to similar setups employing other TOF-PET ASICs, mentioned in the introduction (see \autoref{tab:comparablesetups} for a comprehensive overview). %
CRT values below \SI{300}{\ps} seem to be feasible on system level using the TOFPET2 ASIC. \cite{rolo2012TOFPETASIC,chen2014DedicatedReadoutASIC,bugalho2017ExperimentalResultsTOFPET2}.

Only the lowest voltages applied revealed an influence of the $V_\mathsf{th\_T1}$ discriminator threshold on the CRT due to the reduced photon detection efficiency of SiPMs at lower overvoltages.
A reduced photon detection efficiency results in slower slopes of the SiPM signal since less photons are detected by the SiPM.
Furthermore, the jitter of the point in time when the first photon and the subsequent photons are detected by the SiPM is larger.
Setting a lower trigger threshold therefore leads to a better time resolution, as this requires the SiPM to detect less photons, which leads to the timestamp of the event being generated at an earlier point in time with a smaller jitter than for higher thresholds and less photons being detected by the SiPM.
This is known as the time walk effect \cite{knoll2010radiation}.
The non-existing performance difference for the investigated values of $\mathsf{vth\_t1}$ over a wide range of higher overvoltages suggests that the photon flux at a very early time point is high for all these overvoltages and the SiPM signals rise quickly and reach the set discriminator thresholds.
The general time resolution degradation, which can be noticed for overvoltages over \SI{6}{V}, is probably related to an increased noise contribution with rising overvoltage.

Lower temperatures are beneficial for the performance.
The dark count rate of the SiPM will increase with rising temperature.
To clearly attribute the observed superior performance at low temperatures to either the SiPM or the ASIC or both, one would need to control the temperature for both components separately which is not possible with the current setup.

The timing effects of the trigger scheme resulting in the appearance of satellite peaks in the coincidence time difference spectrum have not been reported so far.
We could show that the effect is clearly related to $\mathsf{delay\_T1}$ element and not a erroneous timeshift by one clock cycle which would very likely be observed at $\pm\SI{5}{\ns}$.
Our working hypothesis is that if the trigger circuit (\autoref{sec:triggergeneration} and \autoref{fig:fig03}) operates as expected, it allows to reject noise events with a signal height smaller than $V_\mathsf{th\_T2}$ but still preserves the timestamp precision of the lower $V_\mathsf{th\_T1}$ threshold.
In this nominal operation case, the time stamp is generated by the delayed signal $\mathsf{do\_T1'}$ which is $t_\mathsf{T1'} = t_\mathsf{T1} + t_\mathsf{delay\_T1}$.
If there is too much noise on $V_\mathsf{out\_T}$ which is higher than $V_\mathsf{th\_T1}$, the $\mathsf{do\_T1}$ and $\mathsf{do\_T1'}$ signals might be active due to noise for a significant amount of time.
If this is the case and a signal pulse from a scintillation event surpasses the voltage level $V_\mathsf{th\_T2}$ in a time window in which $\mathsf{do\_T1'}$ is already enabled, the $\mathsf{do\_T2}$ arrival at the $\mathsf{AND}$ gate will immediately generate the $\mathsf{trigger\_T}$ signal.
Compared to the nominal trigger generation, the timestamp will not incorporate the $t_\mathsf{delay\_T1}$ time span.
If there is a mixture of events that are generated nominally and events that arrive during a noise-induced activation of $\mathsf{do\_T1'}$, the relative time difference between those events will be $\Delta t = t_\mathsf{delay\_T1} - (t_\mathsf{T2} - t_\mathsf{T1})$.
$\Delta t$ has a fixed component depending on an ASIC configuration parameter $t_\mathsf{delay\_T1}$ and a variable component $(t_\mathsf{T2} - t_\mathsf{T1})$ which depends in the slope of the voltage signal and the discriminator levels $V_\mathsf{th\_T1}$ and $V_\mathsf{th\_T2}$.
Both detectors can be harmed by this effect individually. 
The time difference is always calculated by subtracting the timestamp of the second detector from the timestamp of the first detector. 
Depending on which detector triggered in the normal mode and which triggered on the second threshold, either a positive or negative additional time difference $\Delta t$ is observed.

The baseline in the satellite peaks calculation for low overvoltages is caused by the background of statistical random coincidences.
Increasing the overvoltage increases the dark count rate and lowering the discriminator threshold $V_\mathsf{th\_T1}$ makes it more prone to be activated by small noise signals, ultimately leading to a higher number of events with a timestamp defined by $t_\mathsf{T2}$.

The slight increase of the peak position difference with rising overvoltage (\autoref{fig:real_shift_vs_ov}) can be explained by the decrease of the time difference between the two discriminator timestamps $t_\mathsf{T1}$ and $t_\mathsf{T2}$, the dynamic part of the observed time difference $\Delta t$.
A higher overvoltage results in a higher gain and a steeper signal slope which results in a smaller time difference between $t_\mathsf{T1}$ and $t_\mathsf{T2}$ and thus in a higher $\Delta t = t_\mathsf{delay\_T1} - (t_\mathsf{T2} - t_\mathsf{T1})$.

The operation point of the TOFPET2 ASIC discriminator thresholds should be chosen carefully for a stable and undeteriorated performance.
Especially the $\mathsf{delay\_T1}$ discriminator should be configured carefully and set to a safe, high-enough value in order to prevent satellite peaks from appearing.
Wrong time stamps could lead to an increased loss of coincidences, an enhanced random rate or wrong TOF information.

\section{Outlook}

Next steps will be to develop a protocol to find stable and suitable threshold settings in order to avoid the appearance of satellite peaks in the coincidence time difference spectra and thus optimize the measurement settings and ASIC performance.
This will most-probably mean to scan the trigger thresholds independently and measure the noise and photonelectron pedestals and to decide on which photoelectron to trigger.
We will investigate if a certain dark count-rate per channel should not be surpassed in order to suppress the wrong timestamp generation.

The energy-linearization model will be checked for baseline effects and non-linearities that are introduced by the digitization using gamma lines of multiple gamma emitters.

Various different single-channel SiPMs as well as SiPM arrays of different vendors will be tested in the future using a similiar evaluation protocol as presented in this work.
This will lead to operating multiple channels of the ASIC simultaneously and to change the scintillator geometry to a design that could actually be used to build up a PET detector block.
The required increase in scintillator thickness is expected to deteriorate the timing resolution.
A decline in the energy resolution is expected for larger arrays of scintillator elements coupled to a sensor board built from multiple SiPMs.

If the TOFPET2 ASIC proves to be a viable candidate to build larger systems with, we plan to design a MR-compatible sensor board and test the ASIC under MRI conditions employing similar interference-test protocols as we used for the evaluation of our previously evaluated PET/MRI technology which employs digital SiPMs \cite{weissler2015hyperionIID, wehner2015PETMRInterference, wehner2014petmriinterference, schug2015tofring}.

\section{Acknowledgments}

We thank Ricardo Bugalho and Luis Ferramacho from PETsys for providing support and answering our many questions.
We thank Florian Schneider from KETEK for sharing an analysis script that helped to get started with our analysis framework.

\ifCLASSOPTIONcaptionsoff
  \newpage
\fi

\printbibliography[heading=bibnumbered]

\end{document}